\documentclass[aps,pra,10pt,twocolumn,superscriptaddress,showpacs,preprintnumbers,amsmath,amssymb,longbibliography]{revtex4-1}

\usepackage{color}
\usepackage{graphicx}
\usepackage{bm}
\usepackage{dcolumn}
\usepackage{verbatim}
\usepackage{mathbbol}
\usepackage{makeidx}
\usepackage{amsmath}
\usepackage{placeins}
\usepackage{simplewick}
\definecolor{dark_red}{rgb}{0.75,0,0}
\definecolor{dark_purple}{rgb}{0.75,0,0.75}
\definecolor{dark_blue}{rgb}{0,0,0.75}
\definecolor{dark_green}{rgb}{0,0.60,0}
\usepackage[colorlinks,linkcolor=blue,citecolor=blue,linkcolor=blue]{hyperref}

\usepackage{array,kantlipsum}
\usepackage[dvipsnames]{xcolor}
\usepackage{tikz}
\usetikzlibrary{fit}
\usetikzlibrary{
  shapes.multipart,
  matrix,
  positioning,
  shapes.callouts,
  shapes.arrows,
  calc}
  
  \definecolor{myyellow}{RGB}{245,177,0}
\definecolor{mysalmon}{RGB}{255,145,73}

\usepackage{smartdiagram}
\usesmartdiagramlibrary{additions}
\usetikzlibrary{babel}
\tikzstyle{bag} = [align=center]

\usepackage[T1]{fontenc}
\usepackage{courier}
\usepackage{xcolor}
\usepackage{lmodern}
\usepackage{listings}
\usepackage[margin=1.0in]{geometry}
\definecolor{mygreen}{rgb}{0,0.6,0}
\definecolor{mygray}{rgb}{0.5,0.5,0.5}
\definecolor{mymauve}{rgb}{0.58,0,0.82}
\begin{document}

\def \scale {0.4}
\def \scaletwo {0.15}
\newcommand{\dr}[1]{\textcolor{dark_red}{#1}}
\newcommand{\dpu}[1]{\textcolor{dark_purple}{#1}}
\newcommand{\db}[1]{\textcolor{dark_blue}{#1}}
\newcommand{\orr}[1]{\textcolor{blue}{\bf[#1]}}
\newcommand{\bpsi}{\boldsymbol{\psi}}
\newcommand{\sumint}{\sum\hspace{-12pt}\int}
\newcommand{\h}[1]{\phantom{#1}}
\newcommand{\vsep}{\vspace{0.4cm}\noindent}
\newcommand{\bws}[0]{\hspace{-1mm}}
\newcommand{\pnt}[1]{{\bf\dr{$\bullet$\,\,#1\,\,$\bullet$\,\,}}}

\newcommand{\SCF}{{\scriptscriptstyle\mathrm{SCF}}}
\newcommand{\SCFp}{{\scriptscriptstyle\mathrm{SCF}^+}}
\newcommand{\XUV}{{\scriptscriptstyle\mathrm{XUV}}}
\newcommand{\IR}{{\scriptscriptstyle\mathrm{IR}}}
\newcommand{\HH}{{\scriptscriptstyle\mathrm{HH}}}
\newcommand{\RABITT}{{\footnotesize\textsc{RABITT }}}
\newcommand{\APT}{{\scriptscriptstyle\mathrm{APT}}}
\newcommand{\CAP}{{\scriptscriptstyle\mathrm{CAP}}}

\newcommand{\XCHEM}{{\textsc{xchem}}}
\newcommand{\ASTRA}{{\textsc{astra}}}
\newcommand{\NEWSTOCK}{{\textsc{newstock}}}
\newcommand{\TAEC}{{\textsc{taec}}}
\newcommand{\CKMESA}{{\textsc{ck-mesa}}}
\newcommand{\LUCIA}{{\textsc{lucia}}}
\newcommand{\DALTON}{{\textsc{dalton}}}
\newcommand{\MOLCAS}{{\textsc{molcas}}}
\newcommand{\GBTOlib}{{\textsc{gbto}{\footnotesize lib}}}
\newcommand{\UKRmol}{{\textsc{ukr}{\footnotesize mol}}}
\newcommand{\UKRmolPlus}{{\textsc{ukr}{\footnotesize mol+}}}

\newcommand{\varA}[1]{{\operatorname{#1}}}

\newcommand{\sjs}[6]{
   \left\{\hspace{-1mm}
       \begin{array}{ccc}
         #1 & \hspace{-1mm}#2 &\hspace{-1mm} #3 \\
          #4 & \hspace{-1mm}#5 &\hspace{-1mm} #6 
       \end{array}\hspace{-1mm}
   \right\}
}

\title{ASTRA: a Transition-Density-Matrix Approach to Molecular Ionization}

\newcommand{\ucf}{Department of Physics \& CREOL, University of Central Florida, Orlando, Florida 32816, USA}
\newcommand{\conicet}{Centro Atómico Bariloche, CNEA, CONICET, and Instituto Balseiro, UNCuyo, 8400 Bariloche, Argentina}
\newcommand{\aarhus}{Department of Chemistry, Aarhus University, Langelandsgade 140, DK 8000 Aarhus C, Denmark}
\newcommand{\nist}{Applied and Computational Math Division, NIST, 100 Bureau Drive, Gaithersburg, MD, 20899, USA}

\author{Juan M. Randazzo}
\affiliation{\ucf}
\affiliation{\conicet}

\author{Carlos Marante}
\affiliation{\ucf}

\author{Siddhartha Chattopadhyay}
\affiliation{\ucf}

\author{Barry I. Schneider}
\affiliation{\nist}

\author{Jeppe Olsen}
\affiliation{\aarhus}

\author{Luca Argenti}\email{luca.argenti@ucf.edu}
\affiliation{\ucf}

\date{\today}

\begin{abstract}
We describe \ASTRA{} (AttoSecond TRAnsitions), a new close-coupling approach to molecular ionization that uses many-body transition density matrices between ionic states with arbitrary spin and symmetry, in combination with hybrid integrals between Gaussian and numerical orbitals, to efficiently evaluate photoionization observables. Within the transition-density-matrix approach, the evaluation of inter-channel coupling is exact and does not depend on the size of the configuration-interaction space of the ions. Thanks to these two crucial features, \ASTRA{} opens the way to studying highly correlated and comparatively large targets at a manageable computational cost. Here, \ASTRA{} is used to predict the parameters of bound and autoionizing  states of the boron atom and of the N$_2$ molecule, as well as the total photoionization cross section of boron, N$_2$, and formaldehyde, H$_2$CO. Our results are in excellent agreement with theoretical and experimental values from the literature. 
%As a proof of principle of \ASTRA{}'s ability to tackle larger targets, we report preliminary results for the photoionization cross section of magnesium-porphyrin (MgH$_{12}$C$_{20}$N$_4$), a biologically relevant  metallorganic complex with as many as 37 atoms.
\end{abstract}

\maketitle

\section{Introduction}\label{sec:intro}

Excitation of correlated electronic motion is at the core of light-induced chemical transformations in matter~\cite{KrausAngewandte2018}. The advent of attosecond spectroscopies~\cite{Hentschel2001,Paul2001b,Krausz2009} has made it possible to study electron dynamics on its natural sub-femtosecond timescale~\cite{Nisoli2017,Palacios2020,Borrego-Varillas2022,LeoneScience2023}. Continuous advances in the XUV and soft-x-ray ultrafast technologies, pursued at large free-electron-laser facilities~\cite{Feldhaus2013,Hartmann2018} as well as in several attosecond laboratories around the world~\cite{Popmintchev2010,Sansone2011,Chini2014,Calegari2016,Kuhn2017,Midorikawa2022}, have continued extending the range of attosecond-pulse parameters to still shorter durations, larger intensities, energies, and repetition rates~\cite{Saito2019,Duris2020,Callegari2021,Mirian2021}. 
At the same time, the focus of attosecond experiments has moved from atomic systems and small molecules to larger systems with chemical and biological relevance~\cite{Calegari2014,Ueda2019,Mansson2021}.

Attosecond pump-probe observables, such as coincidence photoelectron distributions and transient absorption spectra, are complex functions of the laser parameters that reflect the correlated and entangled dynamics of the probed system only indirectly. Due this complexity, \emph{ab initio} methods beyond the single-particle approximation are essential to reconstruct and ultimately control the many-body motion excited by short light pulses~\cite{Armstrong2021}.

Here we present \ASTRA{} (AttoSecond TRAnsitions), a wave-function close-coupling approach based on a new transition-density-matrix close-coupling (TDMCC) formalism, suited to describe \emph{ab initio} the ionization processes that are made accessible by the use of new sources of coherent ultrashort light. \ASTRA{} is designed to describe both atoms and larger molecules, for both low- and high-energy photoelectrons. It takes exchange and correlation into account to the full extent allowed by state-of-the-art large-scale configuration-interaction quantum-chemistry methods~\cite{Olsen2000a}. \ASTRA{} admits a natural extension to double-ionization processes, which are needed to observe the motion of correlated electron pairs in real time, as well as to multiple disjoint molecular fragments that share electrons. Thanks to these properties, \ASTRA{} promises to appreciably expand the capabilities made available by existing molecular ionization codes.

This paper is organized as follows. Sec.~\ref{sec:Context} gives a short overview of the theoretical methods currently available or under active development to describe the ionization of polyelectronic molecular systems, and where \ASTRA{} fits in this context. 
Sec.~\ref{sec:Method} defines the close-coupling (CC) states~\cite{Burke1962}, the expressions for arbitrary operators within the TDMCC formalism, and relevant observables. Sec.~\ref{sec:Implementation} illustrates how TDMCC is numerically implemented in \ASTRA{}. Sec.~\ref{sec:Results} presents \ASTRA{} results for the boron atom, the nitrogen molecule, and formaldehyde. %, and it shows a proof-of-principle calculation for magnesium-porphyrin. 
Sec.~\ref{sec:Conclusions} offers conclusions and perspectives. Appendix~\ref{app:SecondQuantization} summarizes second-quantization formalism. Appendix~\ref{app:TDM} introduces Transition Density Matrices and their spin-adapted reduced counterparts. Appendix~\ref{app:MatrixElements} details the derivation of the matrix elements between spin-adapted close-coupling states. Finally, App.~\ref{app:CIS} summarizes the CI-singles approximation.

\section{Context~\label{sec:Context}}

A broad group of many-body methods for atomic and molecular ionization assumes a single-reference for both the initial bound state and the final state in the continuum. These include: the \textsc{xatom}~\cite{Rudek2012} and \textsc{xmolecule} codes~\cite{Hao2019}, suited for estimating fragment yields following multiple sequential ionization steps by intense x-ray Free-Electron Lasers (XFEL) pulses~\cite{Rudenko2017}; the \textsc{e-Polyscat} code~\cite{Natalense1999,Gong2022}, which has been applied to single-photon and multiphoton single ionization in the static-exchange approximation; \textsc{tiresia} B-spline density-function method, based on single center~\cite{Venuti1998,Stener1998} and polycentric expansion~\cite{Toffoli2002,Decleva2022},  which was applied to strong-field ionization~\cite{Petretti2010},  core photoionization~\cite{Canton2011,Plesiat2012b,Argenti2012}, pump-probe ionization of molecules with biological relevance~\cite{Calegari2014,Lara-Astiaso2018}, and high-harmonic generation~\cite{Neufeld2019}; \textsc{octopus}, a general-purpose time-dependent density-function theory (TDDFT) grid method applicable also to attosecond photoelectron spectroscopy~\cite{DeGiovannini2012} and transient-absorption spectroscopy~\cite{DeGiovannini2013a}.
While single-reference methods reproduce the essential features of strong-field ionization processes~\cite{Awasthi2008} of core photoelectron spectroscopy~\cite{Kukk2013} and of processes that lead to the complete fragmentation of a target~\cite{Ayuso2014}, wave-function-based multi-reference approaches are needed for those cases in which multiply excited states and the entanglement between photo-fragments play a fundamental role.

Time-dependent (TD) multi-configuration (MC) self-consistent-field (SCF) approaches offer a compact description for the wave function. These include multi-configuration TD Hartree~\cite{Lode2020} and Hartree-Fock~\cite{Haxton2011,Sawada2016,Liao2017}, 
TD restricted-active-space (RAS) SCF~\cite{Miyagi2013,Omiste2021},
TD generalized-active-space (GAS) SCF~\cite{Bauch2014,Chattopadhyay2015a},
TD complete-active-space (CAS) SCF~\cite{Sato2013a},
TDMCSCF~\cite{LiSato2021}, and TD coupled-cluster methods~\cite{SatoCC2018}.
These methods are effective for optical observables, whereas photoelectron observables are more difficult to converge since parent ions evolve even in the asymptotic limit, in absence of fields.

The remaining methods rely on the close-coupling approach (CC), which guarantees the correct asymptotic behavior of the wave function, and they give rise to a stable linear evolution. These methods differ in the level of accuracy with which the ions are represented, in the approximations made to evaluate the matrix elements between CC states, and in the limits imposed to the size and energy of the system. 

In its simplest configuration-interaction singles (CIS) formulation~\cite{Toffoli2016,Hoerner2020,Carlstrom2022}, 
the CC space is described by single excitations from a reference single-determinantal function. 

More general close-coupling approaches include correlation in the $N-1$-electron parent ions, as well as localized $N$-electron configurations to reproduce the short-range correlation of the neutral system in either bound or continuum states. Furthermore, different methods adopt different ways of reconciling the convenience of Gaussian-type orbitals (GTO) to describe bound states with the need of employing numerical bases to represent diffuse Rydberg and continuum orbitals, such as monocentric~\cite{Bachau2001} and polycentric~\cite{Toffoli2002} B-splines, or monocentric~\cite{Yip2008} 
and polycentric FEDVR functions~\cite{Greenman2017}.

The R-matrix method for molecular ionization~\cite{Burke1977}, originally implemented for electron scattering and stationary regimes (\UKRmol{})~\cite{Gorfinkiel2005}, 
has recently been extended to time-dependent photoionization processes (\UKRmolPlus{})~\cite{Harvey2014,Masin2020,Brown2020}. 
\UKRmolPlus{} employs hybrid Gaussian B-spline bases, and it can deal with arbitrary hybrid integrals.
The K-matrix RPA approach has been applied to one- and two-photon transitions in stationary regime~\cite{Cacelli1991,Cacelli2001}.%,Moccia2003}.

The Complex-Kohn approach, based on the \textsc{mesa} program, (\CKMESA{}) has the unique feature of representing the effect of short-range correlation by means of an optical potential from a large CI space, which leads to accurate electron-scattering and one-photon photoionization amplitudes~\cite{Schneider1988,McCurdy1989,Schneider1990,Rescigno1993,Williams2012}. 
Recently, \CKMESA{} has been extended to two-photon transitions and applied to resonant pump-probe process~\cite{Douguet2018a}. 
The time-dependent Recursive indeXing code (\textsc{t-Recx}) is a general software for the \emph{ab initio} solution of large-scale time-dependent problems in atomic and molecular physics, including close coupling in a hybrid basis~\cite{Majety2015c,Scrinzi2022}.

The \XCHEM{} code  
circumvents the need of computing some exchange terms in the close-coupling Hamiltonian and some hybrid bielectronic integrals by confining the molecular ion in a spherical region in which all electronic states are expanded in terms of GTOs~\cite{MarantePRA2017,MaranteJCTC2017,KlinkerJPCL2018,Klinker2018,MarggiPoullain2019,Borras2021,Borras2023a}. 
The polycentric B-spline basis developed for the \textsc{tiresia} code has recently been co-opted for the development of a novel molecular ionization code based on the algebraic diagrammatic construction (ADC) to represent correlated many-body states~\cite{RubertiDecleva2018,Ruberti2019a}.
The Multi-Channel Schwinger Configuration Interaction Method (\textsc{mcsci}), an established method for molecular photoionization, based on the graphical unitary group approach~\cite{Bandarage1989,Stratmann1995}, 
has also been applied to the description of time-resolved photoelectron spectroscopy in linear molecules~\cite{Lucchese2007b}.

\ASTRA{} belongs to the latter group of most general CC methods. It differs from the methods above because it relies on a TDMCC formalism. In \ASTRA{}, the intricacies of the many-body interactions within the parent ions and between the ions and the photoelectron(s) are confined to the n-point correlation functions of highly correlated ionic states, which are efficiently computed by a large-scale-CI bound-state code. The computational cost for CC observables is independent of the size of the CI space for the ion. Exchange terms are treated exactly, so that the size of the target and the photoelectron energy are not constrained. \ASTRA{} predictions are in excellent agreement with theoretical and experimental benchmarks for atoms and small molecules.

\section{TDMCC formalism}\label{sec:Method}

\subsection{The Close-Coupling Ansatz}
In single photoionization processes, an initially bound molecular system $\mathrm{M}$ breaks into a photoelectron and a bound state $A$ of the $\mathrm{M}^+$ molecular parent ion. In the case of the absorption of a single photon $\gamma$ from the ground state, for example, the process reads
\begin{equation}\label{eq:one-photon-single-photoionization}
\mathrm{M}_g + \gamma \longrightarrow \mathrm{M}^+_A + e^-.
\end{equation}
Once the two charged fragments are far apart, the parent ion can be found in only a finite number of states, i.e., all those permitted by conservation of energy, $E_{\mathrm{A}}-E_g < \hbar \omega_\gamma$, where $\omega_\gamma$ is the photon frequency. When the parent ion and electron are well separated, the configuration space may be represented by a linear combination of parent ions times arbitrary one-particle states for the photoelectron,
\begin{equation}\label{eq:close-coupling-ansatz}
\Psi \simeq \hat{\mathcal{A}}_N{\sum}_{A,P}  c_{A,P} [ \Phi_A(\mathbf{x}_1\cdots \mathbf{x}_{N-1}) \varphi_P(\mathbf{x}_N) ],
\end{equation}
where $\mathcal{A}_N$ is the antisymmetrization operator~\cite{Hamermesh},
$\varphi_P(\mathbf{x}_N)$ is a spin-orbital for the $N$-th electron, $\Phi_{A}(\mathbf{x}_1\cdots \mathbf{x}_{N-1})$ is an ionic state,  $\mathbf{x}=(\vec{r},\zeta)$ indicates an electron's spatial and spin coordinates, and $c_{A,P}\in\mathbb{C}$ are expansion coefficients. The \emph{ansatz}~\eqref{eq:close-coupling-ansatz}, known as close-coupling (CC) approximation~\cite{Burke1962}, accurately represents bound and autoionizing Rydberg series and multi-channel scattering states for large values of the photoelectron radial coordinate. 

At short range, a truncated CC expansion is able only in part to reproduce the correlated dynamics of the parent ion and the photoelectron. The CC approximation can be improved by including $N$-electron short-range configurations, which can be generated by extending the CC expansion to closed channels with thresholds well above the energy of interest.
If the additional closed channels correspond to realistic ionic states, this procedure fails to account for the contribution of the ionic continuum, which may be relevant. A more effective and rapid convergence can be obtained by artificially confining the ionic states to a finite set of localized orbitals. In this paper, we follow this latter approach. 

\subsection{CC Matrix Elements \label{sec:Method:MatrixElements}}
This section defines single-ionization CC states and derives the matrix elements of arbitrary spin-free one-body and two-body operators between them. 
The CC partial-wave channel state formed by an ionic state $A$ and a normalized spin-orbital $P$, to represent the photoelectron, is
\begin{equation}\begin{split}
&\Psi_{A,P}(\mathbf{x}_1,\cdots,\mathbf{x}_{N})=\\
&(-1)^{N-1}\sqrt{N}\hat{\mathcal{A}}_N\Phi_A(\mathbf{x}_1,\cdots,\mathbf{x}_{N-1})\,\phi_P(\mathbf{x}_N).
\end{split}\end{equation}
In this paper, it is convenient to employ the formalism of second-quantization, in which the partial-wave states are expressed as
\begin{equation}\label{eq:SIfq}
\Psi_{A,P}(\mathbf{x}_1,\cdots,\mathbf{x}_{N})\,\,\longleftrightarrow\,\, |A,P\rangle = a_P^\dagger |A\rangle.
\end{equation}
For a summary of the second-quantization formalism used in the present work, see App.~\ref{app:SecondQuantization}.

To determine the bound and scattering electronic states of a molecule, and the electron dynamics induced by the exchange of photons with an external field, it is necessary to evaluate the matrix elements of the Hamiltonian and dipole operators between CC states. In second-quantization formalism, operators are written as combinations of strings of creators and annihilators. One-body operators are (see ch.1 in~\cite{MEST2000}):
\begin{eqnarray}
\hat{O}=o_{RS}\, a^\dagger_{R} a_{S},\label{eq:1BOp}
\end{eqnarray}
where the sum over repeated indexes is implied, and $o_{RS}=\langle R | \hat{o} | S\rangle$ is the operator matrix element between spinorbitals for a single particle.
The electron-electron repulsive Coulomb potential, a two-body operators, is
\begin{eqnarray}
 \hat{G}=\frac{1}{2}\,[PQ|RS]\,a^\dagger_{P}a^\dagger_{R}a_{S}a_{Q}\label{eq:2BOp}
\end{eqnarray}
where $[PQ|RS]$ is the electrostatic repulsion term between the charge distributions $\rho_{PQ}(\vec{r})=\sum_\zeta \phi_P^*(\vec{r},\zeta)\phi_Q(\vec{r},\zeta)$ and $\rho_{RS}(\vec{r})=\sum_\zeta \phi_R^*(\vec{r},\zeta)\phi_S(\vec{r},\zeta)$,
\[
[PQ|RS]=\int d^3r_1\,d^3r_2\,\rho_{PQ}(\vec{r}_1)\,r_{12}^{-1}\,\rho_{RS}(\vec{r}_2).
\]

The calculation of the Hamiltonian and dipole matrix elements between CC states is thus reduced to evaluating strings of up to six creation and annihilation operators between ionic states, e.g.,
\[
\langle A,P | \hat{G} | B,Q\rangle = \frac{1}{2}\,[TV|RS]\,\langle A | a_Pa^\dagger_{T}a^\dagger_{R}a_{S}a_{V}a^\dagger_Q | B\rangle.
\]
Using the permutation symmetry of creation and annihilation operators, these operator strings can be written as a linear combination of strings in normal order (creators to the left, annihilators to the right). The matrix elements between ionic states of operators strings in normal order are known as transition density matrices (TDM). The one-, two- and three-body TDMs are identified by the following notation~\cite{McWeeny}
\begin{equation}\label{eqs:TDMs}
\begin{split}
\rho^{BA}_{Q,P}&\equiv \langle A | a^\dagger_P a_Q |B\rangle,\\
\pi^{BA}_{RS,PQ}&\equiv \langle A | a^\dagger_P a^\dagger_Q a_S a_R|B\rangle,\\
\gamma^{BA}_{ STU, PQR}&\equiv \langle A | a^\dagger_P a^\dagger_Q a^\dagger_R a_U a_T a_S|B\rangle.
\end{split}
\end{equation}

As shown in App.~\ref{app:SecondQuantization}, using these definitions it is possible to compute matrix elements between single ionization CC states for different operators by combining the TDMs with the operator matrix elements between basis orbitals.
The overlap between CC states, for example, reads
\begin{equation}\label{eq:Overlap}
\langle A,P|B,Q\rangle = \langle P|Q\rangle \delta_{AB}- \rho^{BA}_{PQ}.
\end{equation}
Even if the overlap between different ionic states is diagonal by construction (the ionic states are eigenstates of the same model Hamiltonian), $\langle A | B\rangle=\delta_{AB}$, the overlap between the orbitals used to augment the ionic states to form the CC states, in general, is not: $\langle P | Q\rangle = s_{pq}\delta_{\pi\theta}$, where $\langle \mathbf{x}|P\rangle = \varphi_{p}(\vec{r})\,^2\chi_\pi(\zeta)$, $\langle \mathbf{x}|Q\rangle = \varphi_{q}(\vec{r})\,^2\chi_\theta(\zeta)$, $s_{pq}=\int d^3r \varphi_p^*(\vec{r})\varphi_q(\vec{r})$. 
The spin orbital $\langle \mathbf{x}|P\rangle$ is defined as the product of a spatial square-integrable function $\varphi_p(\vec{r})\in\mathcal{L}^2(\mathbb{R}^3)$ and the spin wave function $^2\chi_\pi(\zeta)$, where $2$ is the electron multiplicy and $\pi\in\left\{-\frac{1}{2},\frac{1}{2}\right\}$ is the spin projection. Thanks to this generality, the TDMCC formalism is compatible with non-orthogonal numerical orbitals such as B-splines~\cite{Bachau2001} and FEDVR functions~\cite{Rescigno2000}, which are ideally suited to represent Rydberg and photoelectron wavepackets.
The matrix elements of one-body operators contain both one- and two-body TDMs, \begin{eqnarray}\label{eq:1BOp2}
\begin{split}
\langle A,P | \hat{O} |B,Q\rangle =  \delta_{AB}\,o_{PQ}+\langle P|Q\rangle\, O_{AB}\\
-\rho^{BA}_{P,R}\,o_{RQ} - o_{PS}\,\rho^{BA}_{S,Q}+ o_{RS}\,\pi^{BA}_{S P,Q R}
\end{split}
\end{eqnarray}
where $O_{AB} = o_{RS}\,\rho^{BA}_{S,R}$ is a matrix element between ionic states. Finally, two-body operators require TDMs up to the third-order,
\begin{eqnarray}\label{eq:tbocc}
\begin{split}
\langle A,P&|\hat{G}|B,Q\rangle=\langle P |Q\rangle\,G_{AB}+
 \\
&+[PQ|RS]\rho^{BA}_{S,R}-[PS|RQ]\rho^{BA}_{S,R}\\
&+[PT|RS]\pi^{BA}_{TS,RQ} + [QT|RS]\pi^{BA}_{PS,RT}  \\
&-\frac{1}{2}[TU|RS]\gamma^{BA}_{USP,TRQ}
\end{split}
\end{eqnarray}
where $G_{AB}=\tfrac{1}{2}\left[TU|RS\right]\pi^{BA}_{US,TR}$.

Expressions  (\ref{eq:Overlap}-\ref{eq:tbocc}) are valid for any value of the spin quantum numbers of the excited orbitals and  the ion states. Since in the present work we restrict our attention to spin-free operators, however, it is convenient to determine the matrix elements between spin-adapted CC states (SACC), $|A,p;S\Sigma\rangle$, in which the spin of the ion and of the additional electron are coupled to a well-defined spin $S$ and spin projection $\Sigma$,
\begin{equation}\label{eq:coupledstate}
\begin{split}
|A,p;S\Sigma\rangle&
=\sum_{\Sigma_A\pi}C_{S_A \Sigma_A,\frac{1}{2}\pi}^{S\Sigma} |A_{\Sigma_A},p_\pi\rangle 
=\\
&=\sum_{\Sigma_A\pi}C_{S_A \Sigma_A,\frac{1}{2}\pi}^{S\Sigma} a_{p_\pi}^\dagger|A_{\Sigma_A}\rangle 
\end{split}
\end{equation}
where we have expanded the capital-letter index $P$ into the (lower case) orbital index $p$ and spin projection (Greek letter) $\pi$, $C^{S\Sigma}_{S_{A},\Sigma_{A},\frac{1}{2}\pi}$ are Clebsch-Gordan coefficients~\cite{Varshalovich1988}, with $S_{A}$ and $\Sigma_{A}$ being the ion spin and spin projection quantum numbers, respectively. The matrix element of a spin-free operator between SACC states, of course, vanishes unless the two states have the same spin $S$ and spin projection $\Sigma$. For brevity, therefore, in the following we will adopt the shorthand notation
\begin{equation}\label{eq:shorthand}
\langle A,p|\hat{M}|B,q\rangle\equiv\langle A,p; S\Sigma|\hat{M}|B,q; S\Sigma\rangle.
\end{equation}

The matrix elements between SACC states are derived in App.~\ref{app:SecondQuantization} and summarized below. As explained more in detail in the next section, it is convenient to partition the spatial orbitals in inactive or core (I), active (A), virtual (V), and external (E) orbitals. The TDMs vanish unless all their indexes correspond to inactive or active orbitals. Furthermore, the expressions for the TDMs with inactive indexes can be simplified in such a way that the TDMs need to be evaluated only for active-orbital indexes. In the following, all sums over orbital indexes appearing in TDMs are assumed to be limited to active orbitals only (or to vanish otherwise). The overlap, for example, reads
\begin{eqnarray}\label{eq:scOve}
\begin{split}
\langle A,p|B,q\rangle=\delta_{AB}\,s_{pq}+W^{BA}_{pq},
\end{split}
\end{eqnarray}
where $W^{BA}_{pq}$ is a reduced one-body TDM obtained by coupling the spin projections of $\rho^{BA}_{P,Q}$ (see App.~\ref{app:MatrixElements}). In~\eqref{eq:scOve}, the $W^{BA}_{pq}$ vanishes unless both $p$ and $q$ are active, in which case $s_{pq}=\delta_{pq}$. In all the other cases, only the $\delta_{AB}\,s_{pq}$ term survives. For a generic one-body operator $\hat{O}$,
\begin{eqnarray}
\langle A,p &| \hat{O} |B,q\rangle =\delta_{AB} \,o_{pq}+s_{pq}\,O_{AB}+\langle\hat{O}\rangle_{\mathrm{core}}\,W^{BA}_{pq}+\nonumber\\
&+W^{BA}_{pr}\,o_{rq} + o_{ps}\,W^{BA}_{sq}+o_{rs}\,P^{BA}_{rs,qp}.\label{eq:sc1B}
\end{eqnarray}
See App.~\ref{app:MatrixElements} for the definition of the reduced TDM $P^{BA}_{rs,qp}$ in terms of $\pi^{BA}_{PS,RQ}$.  
The term $\langle\hat{O}\rangle_{\mathrm{core}}=2o_{xx}$ represents the contribution to the observable from the ionic inactive orbitals.
Finally, the complete Hamiltonian matrix element between SACC states comprises both mono- and bi-electronic components and it involves up to three-body TDMs. The Hamiltonian matrix elements between CC states that do not contain any virtual or external orbital are most readily evaluated using the same quantum-chemistry program that computes the TDMs. For this reason, the current implementation of \ASTRA{} does not explicitly use three-body TDMs.
The general matrix element between SACC in which at least one of the two states involves a virtual or external orbital, reads
\begin{equation}\label{eq:Hcc}
\begin{split}
H_{Ap,Bq}&=\langle A,p| \hat{H} |B,q\rangle =\\
&=\delta_{A B} (s_{pq} E_A + \tilde{h}_{pq})+\\
&+W^{BA}_{pr}\tilde{h}_{rq} +\tilde{h}_{pr}W^{BA}_{rq}\\
&+\delta_{S_AS_B}[pq|rs]Q^{BA}_{sr}+[ps|rq] W^{BA}_{sr}+\\
&+[pt|rs] P^{BA}_{ts,qr} + [qt|rs]P^{BA}_{ps,tr}
\end{split}
\end{equation} 
where $E_A$ is the energy of the parent ion $A$, and where the mono-electronic matrix element $h_{pq}$ has been replaced by an effective Hamiltonian $\tilde{h}_{pq}$ that incorporates the Coulomb and exchange terms with the core electrons,
\begin{equation}\label{eq:htilde}
\begin{split}
\tilde{h}_{pq}=h_{pq} +(2[pq|\tilde{r}\tilde{r}]-[p\tilde{r}|\tilde{r}q]).
\end{split}
\end{equation}
The reduced TDM $Q^{BA}_{sr}$ is defined in App.~\ref{app:MatrixElements} (see~\ref{Eq:Qtensor}).

\subsection{Observables}
In this paper, we focus on the parameters of bound states and autoionizing states of selected atomic and molecular systems, as well as on the total photoionization cross section of each system from its ground state. The energy of the bound states and the position and width of the autoionizing states are obtained by diagonalizing the electronic fixed-nuclei Hamiltonian in the SACC basis described above, with outgoing boundary conditions. The radial functions used to build the CC channels are confined to a quantization box $r<R_{\mathrm{box}}$, with typical size of the order of few hundred Bohr radii. Outgoing boundary conditions are enforced by adding to the Hamiltonian a complex absorption potential (CAP)~\cite{Muga2004}, $V_{\textsc{CAP}}$, with support located near the boundary of the quantization box, 
\begin{eqnarray}
\tilde{H}&=&\sum_{i}\left[\frac{p_i^2}{2}-\sum_A\frac{Z_A}{|\vec{r}_i-\vec{R}_A|}\right]+\sum_{i>j}\frac{1}{|\vec{r}_i-\vec{r}_j|}+\nonumber\\
&+&\sum_{A>B}\frac{Z_AZ_B}{|\vec{R}_A-\vec{R}_B|}+V_{\textsc{CAP}}\\
V_{\textsc{CAP}}&=&-i\sum_{\alpha i} b_\alpha\,\theta(r_i-R_{\alpha})\,(r_i-R_\alpha)^2
\end{eqnarray}
where $Z_A$ and $\vec{R}_A$ are the charge and position of nucleus $A$, while $\vec{r}_i$ and $\vec{p}_i$ are the position and momentum operator of the i-th electron. Here, the CAP is parametrized as the sum of several negative imaginary parabolic potentials, $\alpha=1,\,2,\,\ldots,\,N_{\textsc{CAP}}$, each starting at a prescribed radius $R_\alpha$; $\theta(x)$ is the Heaviside step function, and $b_\alpha$ are real positive constants. In dynamical conditions, the CAP prevents artificial reflections of a photoelectron wavepacket from the box boundary. While local multiplicative CAP are not perfect absorbers~\cite{Muga2004}, the reflection coefficient of an electron from either the box boundary or the CAP itself can easily be reduced to machine precision in any energy range of interest by combining multiple parabolic potentials across a sufficiently large radial distance. For the present work, in which we examine total energies within one Hartree from the ionization threshold, we found that a combination of a broad and shallow complex potential ($b_1\simeq 10^{-6}$, $R_1\simeq R_{\mathrm{box}}-200\,\mathrm{a.u.}$), which efficiently absorbs slow electrons without reflecting them, and a narrower and steeper complex potential ($b_2\simeq 10^{-4}$, $R_2\simeq R_{\mathrm{box}}-100\,\mathrm{a.u.}$), which absorbs fast electrons before they get reflected by the box boundary, is sufficient to achieve convergence. 

Let $|\boldsymbol{\phi}\rangle = (|\phi_1\rangle,|\phi_2\rangle,\ldots)$ be the CC basis and $\mathbf{\tilde{H}}=\langle\boldsymbol{\phi}|\tilde{H}|\boldsymbol{\phi}\rangle$ be the representation of the Hamiltonian with CAPs in this basis. The diagonalization of $\mathbf{\tilde{H}}$ can be written as
\begin{equation}
    \mathbf{\tilde{H}} = \mathbf{U}_{R}\mathbf{\tilde{E}}\mathbf{U}_L^\dagger,
\end{equation}
where $\mathbf{\tilde{E}}_{ij}=\delta_{ij} \tilde{E}_i$ is the diagonal matrix of the complex eigenvalues of the projected $\tilde{H}$, whereas $\mathbf{U}_{L/R}$ are the left/right eigenvectors, normalized so that $\mathbf{U}_{L}^\dagger\mathbf{U}_{R}=\mathbf{1}$.
The complex energies thus obtained can be grouped in three types: i)~those below the first ionization threshold, with negligible imaginary part, which correspond to bound states that have negligible amplitude in the CAP region; ii)~sequences of eigenvalues that branch out from each threshold and rapidly acquire large negative imaginary components, which correspond to non-resonant continuum states; iii)~isolated complex eigenvalues $\tilde{E}_i=\bar{E}_i - i\Gamma_{i}/2$, where $\bar{E}_i=\Re e(\tilde{E}_i)$ and $\Gamma_i = -2\Im m(\tilde{E}_i)$, which are largely independent on the choice of the extinction parameter $b_\alpha$ and represent the complex energies of autoionizing states. In the region where the CAP vanishes, $\forall i,\,\alpha,\,\, r_i\leq R_\alpha$, the right eigenstates of $\mathbf{\tilde{H}}$ can be regarded as Siegert states~\cite{Siegert1939,McCurdy1979,Riss1993}.

The spectral resolution of the Hamiltonian in terms of states that satisfy outgoing boundary conditions also allows us to compute the total photoionization cross section, which in length gauge reads
\begin{equation}\label{eq:opticaltheorem}
\begin{split}
\sigma_{\mathrm{tot}}(\omega)&=\frac{4\pi^2\omega}{c}\sum_{\alpha}|\langle\Psi_{\alpha E_g+\omega} |\hat{\epsilon}\cdot\vec{\mu}|g\rangle|^2,
\end{split}
\end{equation}
where $|g\rangle$ is the ground state of the target, $c\simeq 137.035$ is the speed of light in atomic units, $\hat{\epsilon}$ is the light-polarization unit vector, $\hat{\epsilon}^*\cdot\hat{\epsilon}=1$,  $\vec{\mu}=-\sum_{i=1}^{N_e}\vec{r}_i$ is the electronic dipole moment, and $|\Psi_{\alpha E}\rangle$ are a complete set of orthonormal single-ionization scattering states for all channels $\alpha$ open at the energy $E_g+\omega$, $\langle\Psi_{\alpha E}|\Psi_{\beta E'}\rangle=\delta_{\alpha\beta}\delta(E-E')$. 
Indeed, for any localized wavepacket $|\phi\rangle$,
\begin{equation}\begin{split}
&\sum_{\alpha} \left|\langle \Psi_{\alpha E}|\phi\rangle\right|^2 = \sum_{\alpha}\langle \phi|\Psi_{\alpha E}\rangle\langle \Psi_{\alpha E}|\phi\rangle=\\
&=\langle \phi|\delta(E-H)|\phi\rangle = -\frac{1}{\pi}\Im m\left[\langle \phi|G^+(E)|\phi\rangle\right],
\end{split}\end{equation}
where $G^+(E)=(E-H+i0^+)^{-1}$ is the retarded resolvent~\cite{Newton}. The total photoionization cross section, therefore, can be written as ($|\phi\rangle=\hat{\epsilon}\cdot\vec{\mu}|g\rangle$)
\begin{equation}\label{eq:opticaltheorem}
\begin{split}
\sigma_{\mathrm{tot}}(\omega)=-\frac{4\pi\omega}{c}\Im m
\langle g|\,\hat{\epsilon}\cdot\vec{\mu}\,\,G_0^+(E_g+\omega)\,\,\hat{\epsilon}\cdot\vec{\mu}\,|g\rangle,
\end{split}
\end{equation}
which is nothing other than the optical theorem applied to photoionization processes. Since wavepackets of the form $G^+(E)|\phi\rangle$ contain outgoing components only~\cite{Newton,Taylor}, we can use the numerical spectral resolution of $\boldsymbol{\tilde{H}}$ to write down an explicit expression for the retarded resolvent in the CC basis,
\begin{equation}\label{eq:Resolvent}
    \mathbf{G}_0^+(E) = \mathbf{U}_R\frac{1}{E-\mathbf{\tilde{E}}}\mathbf{U}_L^\dagger.
\end{equation}
Equation~\eqref{eq:Resolvent} is accurate as long as the resolvent is applied to wavepackets localized in the region where the CAP vanishes and the result is evaluated in the same region. This is the case for the dipole operator acting on the ground state of the system, and hence
\begin{equation}\label{eq:opticaltheoremCAP}
\begin{split}
\sigma_{\mathrm{tot}}(\omega)=-\frac{4\pi\omega}{c}\Im m\Big[
\boldsymbol{\mu}^\dagger\,\mathbf{G}_0^+(E_g+\omega)\boldsymbol{\mu}\Big],
\end{split}
\end{equation}
where $\boldsymbol{\mu}=\langle\boldsymbol{\phi}|\,\hat{\epsilon}\cdot\vec{\mu}\,|g\rangle$.
This expression bypasses the calculation of scattering states and it can be evaluated at a negligible computational cost. The calculation of multi-channel scattering states is beyond the scope of the present paper and will be examined in future works.

\section{\ASTRA{} Implementation}\label{sec:Implementation}

In this section, we examine some aspects of the numerical implementation of \ASTRA{}. 

To reproduce molecular ionization observables, \ASTRA{} relies on internal components as well as on external tools: \DALTON{}~\cite{DALTON2014}, \LUCIA{}~\cite{Olsen1988}, and \GBTOlib{}~\cite{Masin2020}. \DALTON{} is a general-purpose Quantum-Chemistry program used to compute an initial set of GTOs as well as mono- and bielectronic integrals between them. \LUCIA{} is a molecular electronic-structure program with a focus on efficient large-scale configuration interactions~\cite{Olsen1988}. Starting from an initial set of orbitals and integrals, \LUCIA{} can carry out sophisticated calculations to optimize the initial orbital, compute correlated bound states, as well as the TDMs between them (see~Sec.~\ref{sec:implementation:lucia} for details).
\GBTOlib{} is a public library, part of the \UKRmolPlus{} molecular scattering package, that can compute electronic integrals between hybrid Gaussian/B-spline functions. \ASTRA{} employs also an additional independent set of B-spline spherical functions external to the molecular region and whose support is effectively disjoint from that of any GTO orbitals. 

\begin{figure*}[hbtp!]
  \centering \includegraphics[width=\textwidth]{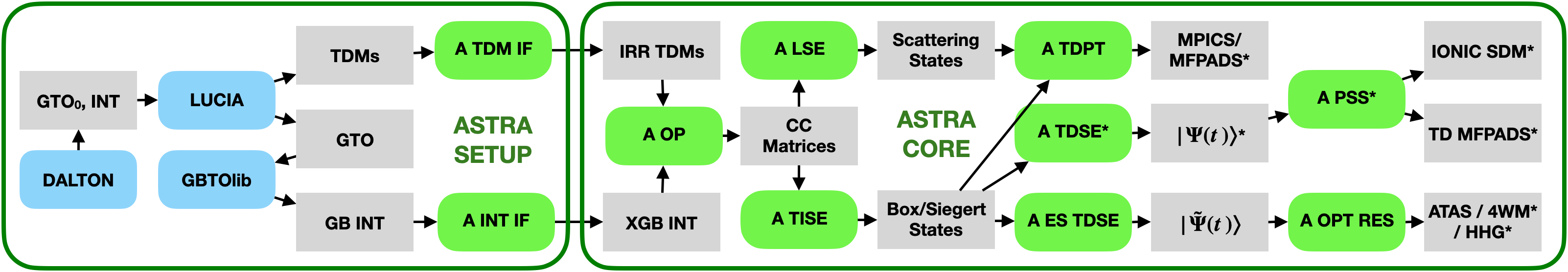}
  \caption{\label{fig:ASTRAflow} \footnotesize \ASTRA{} flow scheme. Rounded boxes represent executables, whereas square boxes represent databases. The ``A'' prefix indicates an executable from the \ASTRA{} suite. In the setup section (left frame), from a set of configuration files (not shown), the \ASTRA{} Setup script invokes, in sequence: i) \DALTON{}, to compute a starting set of GTO and their associated integrals; ii) \LUCIA{}, to optimize the orbitals, to compute the correlated ionic states, as well as the TDM between them; iii) \GBTOlib{}, to compute, from the optimized orbitals, the hybrid integrals with the B-splines in the molecular region. See text for details on the steps that involve only \ASTRA{} components. Starred elements are currently under development.}
\end{figure*}
The flow diagram in Fig.~\ref{fig:ASTRAflow} illustrates how the different external and internal components of \ASTRA{} are connected with each other. The external components are managed by a setup script. One interface (A TDM IF) converts the TDMs generated by \LUCIA{} to their reduced spin-adapted form. A second interface (A INT IF) converts the hybrid integral generated by \GBTOlib{} to the internal database of \ASTRA{} and complements the hybrid integrals with those involving the additional set of B-splines that extends beyond the molecular region. 

The right frame of the flow diagram delimits the core section of \ASTRA{}. The \ASTRA{} operator program (A OP) generates the representation of arbitrary operators in the SACC space, starting from the reduced TDM, the hybrid integrals, and the specifics of the close-coupling configuration file (not shown). Subsequently, the \ASTRA{} TISE program determines the eigenstates and eigenvalues of the system with either vanishing or outgoing boundary conditions. In parallel, the Lippmann-Schwinger Equation solver (A LSE) determines the scattering states. 
The generalized and the confined eigenstates of the system can then be used to evaluate stationary observables, such as the one-photon total ionization cross section (see Sec.~\ref{sec:Results}), the photoelectron emission due to external pulses (A TDSE, A PSS), by solving the TDSE in the full CC basis, or the optical response by solving the TDSE in a reduced model space of Siegert states (A ES TDSE, A OPT RES). 
Due to its reliance on quantum-chemistry external tools, \ASTRA{} only handles molecular spatial symmetry groups that are subgroups of $D_{2h}$~\cite{Hamermesh}.

\subsection{CASCI ionic states and TDM \label{sec:implementation:lucia}}
The molecular orbitals can be determined using \LUCIA{} for several wave function methods: Hartree-Fock,  Complete Active Space Self-Consistent Field calculations (CASSCF), 
and Generalized and Restricted-Active-Space SCF calculations (GASSCF and RASSCF). Molecular ionization processes start from a neutral state and lead to the production of multiple ionic states with different symmetries and multiplicities. It is therefore advantageous to optimize the orbitals so they provide a balanced representation of a set of target states with different number of electrons, multiplicities, spatial symmetry, and energy. Such optimizations go under the name of state-averaged (SA) MCSCF calculations, and \LUCIA{} is capable of optimizing the orbitals for an ensemble of such states. This makes \LUCIA{} a well-suited QC engine for generating the orbitals and CI-expansions that subsequently are used in the  photoionization codes. 

In a subsequent step, \LUCIA{} computes the required TDMs between all the ionic states in the CC expansion, as well as the overlap, Hamiltonian, and dipole matrix elements between all the states obtained by augmenting these ions by any of the active orbitals. For one-electron photoionization processes, one- and two-electron density matrices are sufficient, but \LUCIA{} can also calculate the three-electron TDMs required for double photoionization. The  TDMs are obtained in \LUCIA{} over spin-orbitals and are subsequently spin-adapted by the \ASTRA{} interface. 
The \LUCIA{} code determines the TDMs from CI expansions in the Slater determinant (SD) basis. However, in order to assure that the involved states have well-defined spin of the involved states, these states are defined in terms of  spin-adapted configuration state functions (CSFs). The calculations of the TDMs and the inner part of a direct CI-step are thus preceded by a transformation from the CSF to the SD 
basis.

The expansion of a CI-state in terms of Slater determinants is realized using the formalism of 
spin-strings~
\cite{Knowles1984,Olsen1988,MEST2000}. In this formalism, a SD is expressed as an alpha-spin-string, which  is an ordered product of the creation-operators for the alpha-spin-orbitals, times a beta-spin-string, which is an ordered 
product of creation operators for 
beta-spin orbitals. The use of spin-strings provides a formally simple approach that can perform the CI-optimization and the subsequent calculation of TDMs in an efficient manner. An advantage of the spin-string formalism is that the information required to determine the action of a general operator on a CI expansion is easy to determine and store. In the present implementation, only information about single-electron removal and addition from strings is determined and stored. The action of operators containing several creation or annihilation operators is constructed on the fly from the lists of one-electron removals and additions.

The TDMs are also written and computed in terms of spin-strings~\cite{Olsen2000,Kaehler2017}. A TDM is thus written as matrix
with indices defining spin-strings, 
\begin{equation}\label{eq:TDMSassttings}
\rho^{BA}_{\nu,\mu} = \langle A |\mathcal{O}^\dagger_\mu \mathcal{O}_\nu |B\rangle.
\end{equation}
By letting  $\mathcal{O}$~ be strings of one-, two-, and three-electron annihilation operators,
the one-, two- and three-electron density matrices of  Eq.(\ref{eqs:TDMs}) are obtained.
The various forms of the TDMs are calculated using a common suite of routines, where the states
$ \mathcal{O}_\mu |A\rangle, \mathcal{O}_\nu |B\rangle$~first are determined as expansions in SDs, $|I\rangle$,
\begin{equation}\label{eqs:effstates}
\mathcal{O}_\mu |A\rangle = {\sum}_I C^{A}_{\mu I} |I\rangle, \quad
\mathcal{O}_\nu |B\rangle = {\sum}_I C^{B}_{\nu I} |I\rangle.
\end{equation}
For a given state A, the dimension of $ C^{A}_{\mu I}$ and the operation count for its evaluation is approximately $N^n N_{\mathrm{det}}$, where $N$~is the number of active orbitals, $n$~is the order of the TDM, and $N_{\mathrm{det}}$~
is the number of SDs in the expansion. In a second step, the TDM is
obtained by matrix multiplication,
\begin{equation}\label{eqs:tdmasproduct}
\rho^{BA}_{\nu,\mu} =\langle A |\mathcal{O}^\dagger_\mu \mathcal{O}_\nu |B\rangle={\sum}_I  C^{A\ast}_{\mu I}   C^{B}_{\nu I}.
\end{equation}
This matrix multiplication scales as  $N^{2n} N_{\mathrm{det}}$. For an example with 20 active orbitals and 1 million SDs, the operation count for the evaluation of $C^{B}_{\nu I}$ for a two-electron TDM is about $4 \times 10^8$ and the following matrix-multiplication requires about $3 \times 10^{11}$ operations. In the above estimates, simplifications arising from spatial symmetry were neglected. For a molecule with $N_{\mathrm{IRR}}$ irreducible representations, the operation count the the matrix multiplication is reduced by a factor of about $N_{\mathrm{IRR}}^2$. On a single core of a modern CPU, the discussed TDM for a molecule with 4 irreducible representations may thus be evaluated in a few seconds.

\subsection{Orbital basis\label{sec:Implementation:OrbitalBasis}}
\begin{figure}[hbtp!]
\centering \includegraphics[width=\columnwidth]{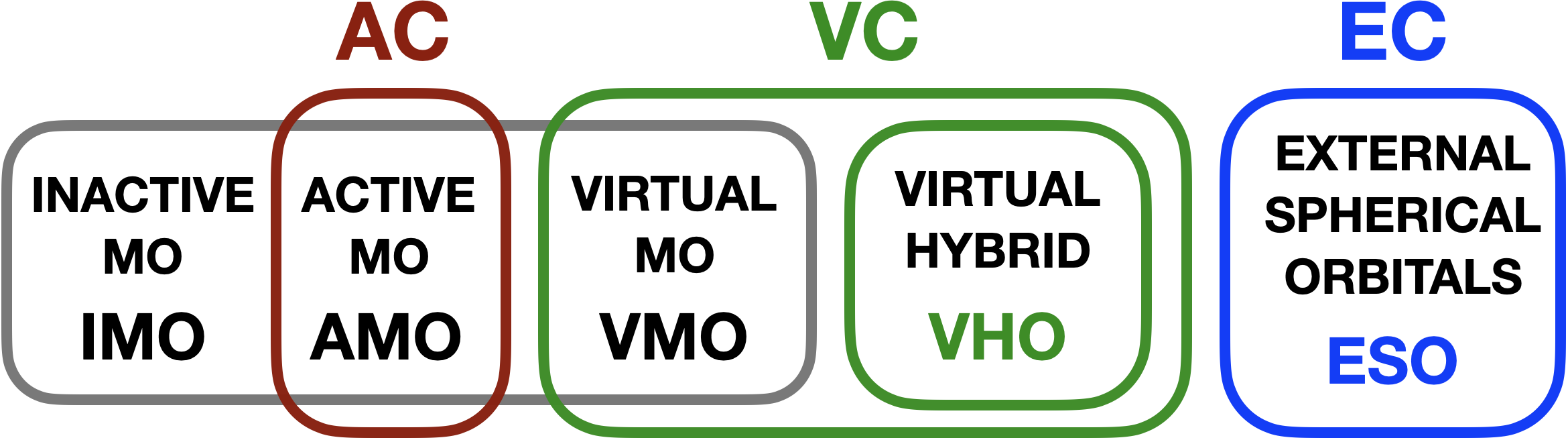}
\caption{\label{Fig:ccs}\footnotesize Schematic classification of the electronic orbitals used in \ASTRA{} CC expansion. The labels over the frames, AC, VC, and EC, indicate the active, virtual, and external component of the close-coupling channel obtained by augmenting a given ion by the full set of the non-inactive orbitals.}
\end{figure}

As commented in Sec.~\ref{sec:Method:MatrixElements}, the expressions for the CC matrix elements suggest a natural partition of the orbitals in inactive, active, virtual, and external, which helps simplify their calculation. Figure~\ref{Fig:ccs} illustrates schematically the different orbitals. 

An initial set of atomic GTO orbitals are used to generate an orthonormal set of molecular orbitals (MO), $\phi^M_i$, $\langle \phi_i^M|\phi_j^M\rangle = \delta_{ij}$, optimized by means of a self-consistent-field calculation (HF, RASSCF, CASSCF, etc.). All MOs are negligible beyond a certain distance $R_{\mathrm{mol}}$ from the molecular barycenter. We call $r<R_{\mathrm{mol}}$ the molecular region. The MOs are divided in inactive orbitals (IMO), which are doubly occupied in all the ions, active orbitals (AMO), which have variable occupation numbers in the ions, and virtual orbitals (VMO), unoccupied in the ions,
\begin{equation}
\left\{\phi^M_{i}\right\} = \left\{\phi^{IM}_{i}\right\}_{i\in I_{IM}} \cup \left\{\phi^{AM}_{i}\right\}_{i\in I_{AM}} \cup \left\{\phi^{VM}_{i}\right\}_{i\in I_{VM}}\nonumber 
\end{equation}
The MOs are assumed to be adequate to represent the correlated state of the $N-1$ electrons of the ion. However, they are normally insufficient to describe the state of an $N-$th electron in interaction with the ion, particularly for energies close to or above the ionization continuum. 
The virtual hybrid orbitals (VHO) are obtained by complementing the virtual MOs with a set of internal spherical B-splines~\cite{Brosolo1992,Bachau2001,Marante2014}, with support within the molecular region,
\begin{equation}
\chi^{\mathrm{int}}_{n\ell m}(\vec{r}) = N_{n}r^{-1}B^{\mathrm{int}}_n(r)\,X_{\ell m}(\hat{r}),
\end{equation}
where $X_{\ell m}(\hat{r})$ are real spherical harmonics and $N_n$ are normalization constants. The internal B-splines are orthonormalized to the MOs,
\begin{eqnarray}
&&\phi^{VH}_i=\sum_{nm\ell} \chi^{\mathrm{int}}_{n\ell m}\,c_{nm\ell,i} + \sum_j \phi^M_j c_{ji},\\
&&\langle \phi^{VH}_i | \phi^M_j\rangle = 0,\quad 
\langle \phi^{VH}_i | \phi^{VH}_j\rangle = \delta_{ij}.
\end{eqnarray}
This second step is carried out by \GBTOlib{}~\cite{Masin2020}, by performing a Gram-Schmidt orthogonalization of the B-spline basis with the MOs (leaving the MOs unchanged) followed by a symmetric orthogonalization of the resulting set. The program also evaluates the monoelectronic and bielectronic integrals involving these hybrid orbitals.

The MOs and VHOs together afford to the CC space considerable flexibility for the description of a photoelectron subject to the polycentric field of the ion. Finally, the external spherical orbitals (ESO) are spherical B-splines that extend the internal B-spline set beyond the molecular region
\begin{equation}
\phi^E_{n\ell m}(\vec{r}) = N_{n}r^{-1}\,B^{\mathrm{ext}}_n(r)\,X_{\ell m}(\hat{r}).
\end{equation}
The MOs and the external B-splines have effectively disjoint support
\begin{equation}
 \phi^{M}_{i}(\vec{r})\,\phi^E_{n\ell m}(\vec{r}) \simeq 0,\quad\forall \vec{r},
\end{equation}
which means that the integral of any local operator between an external B-spline and an hybrid orbital can be expressed in terms of integrals between B-spline functions only.

\subsection{Matrix elements}

\begin{figure}
\centering \includegraphics[width=\columnwidth]{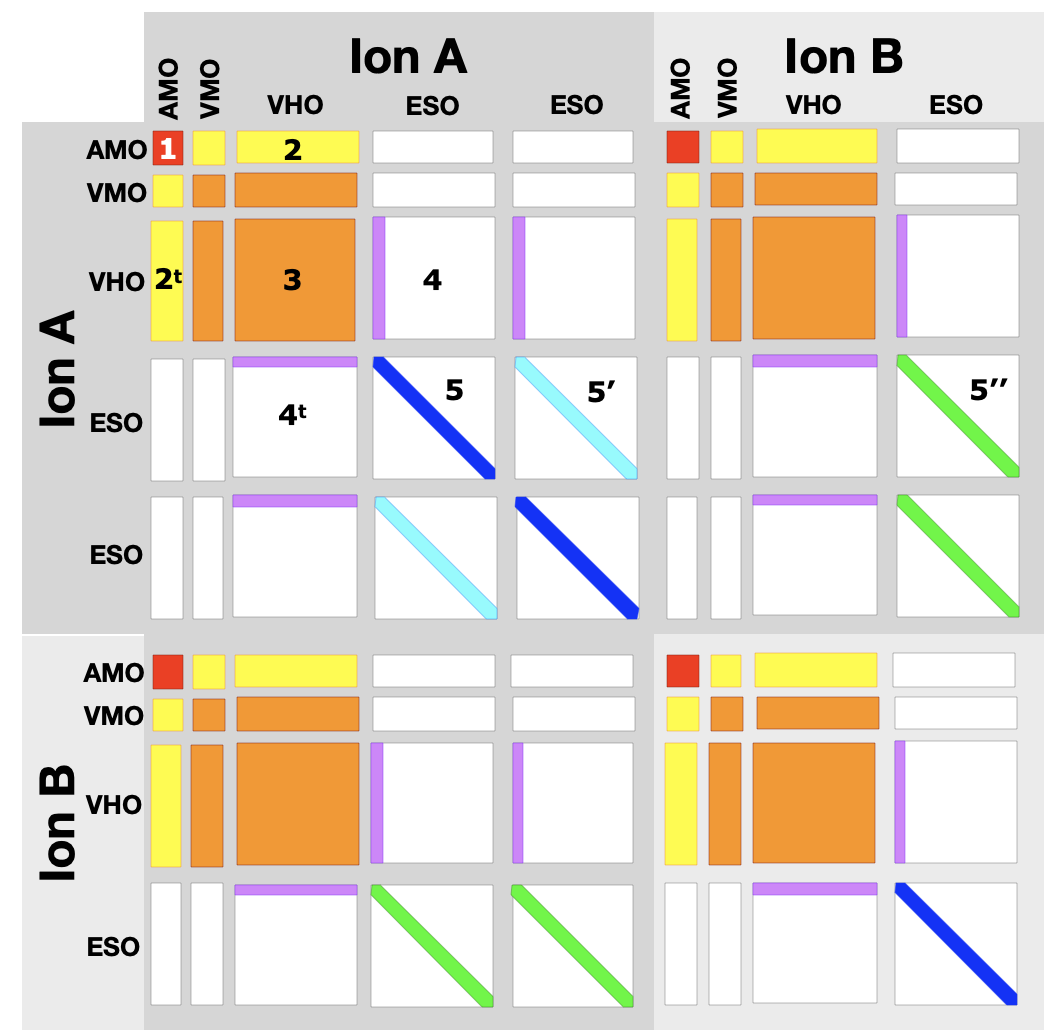}
\caption{\label{Fig:Hmatrix} \footnotesize Block-structure of the Hamiltonian matrix between the CC states. The parts in white are empty. While the blocks are not to scale, it is clear that the matrix is highly sparse. There are essentially five different types of blocks (colored online), which correspond to as many separately optimized formulas and set of integrals: \textbf{1} for~\eqref{eq:Hll}, \textbf{2} for~\eqref{eq:H_AMO_VHO}, \textbf{3} for~\eqref{eq:Hpaqv} and for \textbf{4, 5} see~\eqref{eq:Hpvqv}.}
\end{figure}

Taking into account the different groups of orbitals used to form the CC channels, we can further simplify the expression for the matrix elements of the different operators given in section~\ref{sec:Method}. 
The structure of the Hamiltonian matrix is schematized in Fig.~\eqref{Fig:Hmatrix}, where the colored regions identify coupling between channels while the white areas correspond to regions with null matrix elements. 
The evaluation of the overlap and of one-body operators is relatively simple. The most complex monoelectronic matrix elements to evaluate are those between AC states, which involves both 1B-TDM and 2B-TDM [see~\eqref{eq:sc1B}].  Overall, however, purely monoelectronic operators are straightforward to evaluate. In this section, therefore, we will focus on the Hamiltonian only. Apart for the complex absorbing potential, which affects exclusively the diagonal blocks between external orbitals, the rest of the Hamiltonian is Hermitian. Therefore, without loss of generality, we can limit the description of the Hamiltonian matrix elements to the upper triangular set of blocks of the close-coupling matrix.
The evaluation of an Hamiltonian matrix element requires the 3B-TDM only when both CC states belong to an AC. Since this matrix elements involves MOs only, it is convenient to carry out the calculation directly within \LUCIA{}. \ASTRA{} is then left with the minimal task of evaluating, from the Hamiltonian matrix element between CC states, only those involving at least a VC or EC state, as detailed in App.~\ref{app:MatrixElements:Hamiltonian}.

Let's now consider the general expression~\eqref{eq:Hcc} for the other blocks. For the active-virtual blocks (AV), the Hamiltonian is
\begin{equation}\notag
\begin{split}
H^{AV}_{Ap,Bq}&=\delta_{A B} \tilde{h}_{pq}+W^{BA}_{pr}\tilde{h}_{rq}\\
&+[pq|rs]Q^{BA}_{sr}+[ps|rq] W^{BA}_{sr}+[qt|rs]P^{BA}_{ps,tr},
\end{split}
\end{equation}
whereas the blocks between active and external states are zero, $H^{AE}_{Ap,Bq}=0$. Notice that the AV blocks are the only ones involving 2B-TDMs. Since there are comparatively few such elements, the calculation is inexpensive. For the VV block, the Hamiltonian contains both local and non-local inter-channel coupling potentials,
\begin{equation}\notag
\begin{split}
H^{VV}_{Ap,Bq}&=\delta_{A B} (E_{A}\delta_{pq}+\tilde{h}_{pq})+\\
&+[pq|rs]Q^{BA}_{sr}+[ps|rq] W^{BA}_{sr},
\end{split}
\end{equation}
whereas in the VE and EE blocks only the local potential remains (X\,=\,V,\,E),
\begin{equation}\label{eq:Hpwc}
\begin{split}
H^{XE}_{Ap,Bq}=\delta_{A B} (E_{A}\delta_{pq}+\tilde{h}_{pq})+[pq|rs]Q^{BA}_{sr}.
\end{split}
\end{equation}
The local potential $[pq|rs]Q^{BA}_{sr}$ describes the repulsion between two disjoint distributions of charge, one confined within and one outside the molecular region, i.e., the multipolar electrostatic coupling between channels.
The expression, therefore, can be conveniently rearranged by means of a multipole expansion of the electron-electron repulsion energy
\begin{equation}
\frac{1}{|\vec{r}-\vec{r}'|} = \sum_{\ell m}\frac{4\pi}{2\ell+1}\frac{r_<^\ell}{r_>^{\ell+1}}X_{\ell m}(\hat{r})X_{\ell m}(\hat{r}'),
\end{equation}
where $X_{\ell m}(\hat{r})$ are real spherical harmonics \cite{Homeier1996}. The bielectronic matrix element $[pq|rs]$ then becomes,
\begin{equation}\notag
[pq|rs] = \sum_{\ell m} o^{(\ell m)}_{pq} M^{(\ell m)}_{rs},
\end{equation}
where 
\begin{equation}\notag
o^{(\ell m)}_{pq}=\langle p | \frac{X_{\ell m}(\hat{r})}{r^{\ell+1}}|q\rangle,\quad M ^{(\ell m)}_{rs} = \frac{4\pi\langle r | r^\ell X_{\ell m}(\hat{r}) |s \rangle}{2\ell+1}.
\end{equation}
We can now recognize that the one-body part of the Hamiltonian $\tilde{h}_{pq}$ defined in \eqref{eq:htilde} already contains a multipolar nuclear attraction potential, whose associated set of moments are given by
\begin{equation}
M_{\mathrm{nuc}}^{(\ell m)} = \frac{4\pi}{2\ell+1}\sum_{\alpha}^{N_{\mathrm{at}}} Z_\alpha R_{\alpha}^\ell X_{\ell m}(\hat{R}_\alpha).
\end{equation}
It is convenient, therefore, to gather the nuclear attraction and the electronic repulsion terms in a single transition multipolar momentum $\mathcal{M}^{(\ell m)}_{AB}$, defined as,
\begin{equation}
\mathcal{M}^{(\ell m)}_{AB} = {\sum_{rs}}'M^{(\ell m)}_{rs}Q^{BA}_{sr} + 2 \sum_{x}^{\mathrm{core}}M_{xx}^{(\ell m)} - M_{\mathrm{nuc}}^{(\ell m)},
\end{equation}
The monopolar momentum, of course, is simply the ionic charge, $\mathcal{M}_{AB}^{(00)}=\delta_{AB}Z_{\mathrm{ion}}$, and the expression for the Hamiltonian in the outer region given in \eqref{eq:Hpwc} can therefore be rewritten as:
\begin{equation}\label{eq:HpwcMP}
\begin{split}
H^{XE}_{Ap,Bq}&=\delta_{A B} (s_{pq} E_A + t_{pq} - Z_{\mathrm{ion}} o^{(00)}_{pq})+\\
&+\sum_{\ell m}^{ \ell > 0 } o^{(\ell m)}_{pq} \mathcal{M}^{(\ell m)}_{AB}.
\end{split}
\end{equation}
where $t_{pq}$ is a kinetic-energy matrix element. 

\section{Results}\label{sec:Results}

This section illustrates \ASTRA{} performance by comparing its results with well established and independent single-ionization codes, together with experimental references, when available. 
We apply \ASTRA{} to the calculation of selected bound and resonant state parameters, and total photoionization cross section of three reference model systems: the boron atom, the nitrogen molecule, and formaldehyde, H$_2$CO, as a representative of an open-shell atomic system, a diatomic molecule, and a simple polyatomic molecule, respectively. %, as well as magnesium porphyrin (MgH$_{12}$C$_{20}$N$_{4}$), a large biologically-relevant organo-metallic complex. 
The results are compared with the best values available in the literature, with which we find an excellent agreement. In the case of N$_2$, we consider different levels of approximation to illustrate both the consistency of the method with independent benchmark calculations, such as CIS, as well as the flexibility with which electronic correlation in ionization can be treated already within the current implementation of \ASTRA{}. 
%While only preliminary, the results for magnesium porphyrin illustrate the scalability of the \ASTRA{} approach.  

\subsection{Boron}
In this section, we discuss the computation of the bound state energies, resonance parameters, and photoionization cross-section of atomic boron and compare the results with the \NEWSTOCK{} suite of atomic photoionization codes~\cite{Carette2013} as well as with a dedicated three-active-electron atomic code (\TAEC{})~\cite{Argenti2016}. 
Boron doublet CC space is built from ions with both singlet and triplet multiplicity. This comparison, therefore, is particularly relevant because it allows us to test the correct implementation of the TDMCC method when TDMs between states with different multiplicity are required.
Although both \ASTRA{} and \NEWSTOCK{} use equivalent CC expansions, the two approaches are intrinsically different. \NEWSTOCK{}, as an atomic photoionization code, uses the full $SO(3)$ symmetry of the atom, whereas \ASTRA{} uses the $D_{2h}$ abelian point group.  \NEWSTOCK{} relies on the non-relativistic multi-configuration Hartree-Fock (MCHF) program of the ATSP2K~\cite{FroeseFischer2007} package to compute the parent-ion states. To obtain the equivalent degree of local electron correlation, we employ the same number of ionic states in \ASTRA{} and \NEWSTOCK{} calculations. The ionic states are optimized by including all possible single and double excitations from the reference determinant with active orbitals up to the principal quantum number $n=4$.  \TAEC{}~\cite{Argenti2016} uses a virtually complete two-electron basis for the two active valence electrons of the B$^+$ ion in the field of a polarizable core, as well as an optimized set of several thousands configurations selected from the full-CI three-active-electron space, and a significantly larger CC expansions compared with either \NEWSTOCK{} or \ASTRA{}. As one may expect from a dedicated program, therefore, \TAEC{} results are closer to the experimental values listed in the NIST database~\cite{NIST_ASD}. 
\begin{table}[hbtp!]
\caption{Comparison of the energies of the first 11 singlet and triplet bound states of B II, $E-E_{2s^2}$ (eV).}
\label{tab:boronionen}
\begin{ruledtabular}
\begin{tabular}{lcccc}
	Conf. &  $\ASTRA{}$ & \NEWSTOCK{} &
    \TAEC{}~\cite{Argenti2016} & Exp.~\cite{NIST_ASD} \\
  \hline  
 \rule{0pt}{3ex}$2s 2p$ ($^3{P^o}$) & 4.6563  & 4.6308 & 4.6391  & 4.6317  \\

 $2s 2p$ ($^1{P^o}$) & 9.3240  & 9.3421 & 9.1187  & 9.1000  \\
 $2p^2$  ($^3{P^e}$) & 12.373  & 12.256 & 12.287  & 12.266  \\

 $2p^2$  ($^1{D^e}$) & 12.856  & 12.860 & 12.706 & 12.691   \\
 $2p^2$  ($^1{S^e}$) & 16.012  & 16.001 & 15.835 & 15.827   \\

 $2s3s$  ($^3{S^e}$) & 16.036  & 16.037  & 16.095 & 16.089  \\
 $2s3s$  ($^1{S^e}$) & 17.166  & 17.300  & 17.069 & 17.062  \\

 $2s3p$  ($^3{P^o}$) & 17.839  & 17.831  & 17.858 & 17.853  \\
 $2s3p$  ($^1{P^o}$) & 17.880  & 17.910  & 17.874 & 17.866  \\

 $2s3d$  ($^3{D^e}$) & 18.652  & 18.650  & 18.681 & 18.678  \\
 $2s3d$  ($^1{D^e}$) & 19.230  & 19.380  & 19.183 & 19.178  \\
  
\end{tabular}
\end{ruledtabular}
\end{table}
In constructing the CC expansion for \ASTRA{} and \NEWSTOCK{} calculations, we have included the first 12 parent ions, which are listed in Table~\ref{tab:boronionen}.  For the \ASTRA{} computation, the ionic Hartree-Fock orbitals are generated with \DALTON{}~\cite{DALTON2014} using the aug-cc-pVQZ basis. In \ASTRA{} and \NEWSTOCK{}, the results of the ionic states with dominant configurations up to $2s3d (^1\mathrm{D^e})$ are compared in Table~\ref{tab:boronionen} relative to the $2s^2(^1\mathrm{S^e})$ ionic ground state. The agreement between our results and the experimental values is remarkable. The values computed with \ASTRA{} lie within 0.03 to 0.2 eV of the experimental values, which is deemed sufficient to ascertain the consistency of \ASTRA{}'s implementation.
In both \ASTRA{} and \NEWSTOCK{}, the twelve parent ions are coupled to photoelectrons with angular momentum up to $l_{max}=3$, within a quantization box of 300 a.u., employing B-splines of order 7 and node separation of 0.3~a.u. In \ASTRA{}, $R_\mathrm{mol}=20$ a.u. and a single CAP is used, which starts at $R_\alpha=200$ a.u.
\begin{table}[hbtp!]
\caption{Bound state energies of B I, $E-E_{2s^2 2p}$ (eV).}
\label{tab:boronen}
\begin{ruledtabular}
\begin{tabular}{lcccc}
	Conf. &  $\ASTRA{}$ & \NEWSTOCK{} &
    \TAEC{}~\cite{Argenti2016} & Exp.~\cite{NIST_ASD} \\
  \hline  
    \rule{0pt}{3ex}$2s 2p^2$  ($^4\mathrm{P^e}$)  & 3.584   & 3.564  &         & 3.552   \\
    $2s^2 3s$  ($^2{S^e}$)  & 4.869   & 4.881  & 4.958   & 4.964   \\
    $2s 2p^2$  ($^2{D^e}$)  & 5.916   & 5.927  & 5.939   & 5.933   \\
    $2s^2 3p$  ($^2{P^o}$)  & 5.993   & 5.990  & 6.021   & 6.027  \\
    $2s^2 3d$  ($^2{D^e}$)  & 6.707   & 6.682  & 6.785   & 6.790   \\
\end{tabular}
\end{ruledtabular}
\end{table}

Table~\ref{tab:boronen} lists the first few bound-state energies of B~I with doublet and quartet symmetries. For the $2s^2 2p (^2P^o)$ ground state, the absolute value of energy obtained from \ASTRA{} is -24.600\,14\,a.u. which differs from more accurate non-relativistic MCHF calculations~\cite{FroeseFischer2013} by 0.0538\,a.u. This energy difference is commensurate with the value of 0.044\,735\,a.u estimated for the K-shell correlation energy~\cite{Davidson1991}, which is not accounted for in the present calculation. The residual discrepancy is compatible with the larger active space used in the MCHF calculation.

\begin{table}[hbtp!]
\caption{Comparison of effective quantum number, $n^{*}$ and resonance width $\Gamma$ (a.u.), of $^2{P^o}$ resonances between $2s^2$ and $2s2p(^3{P^o)}$ thresholds. The notation 5.94[-3] stands for $5.94\times10^{-3}$. }
\label{tab:boronrespar2Po}
\begin{ruledtabular}
\begin{tabular}{lcccccc}
               & \multicolumn{2}{c}{\ASTRA{}} & \multicolumn{2}{c}{\NEWSTOCK{}} & \multicolumn{2}{c}{\TAEC{}~\cite{Argenti2016}} \\
  
        Conf.  & $n^{*}$   & $\Gamma$      & $n^{*}$    & $\Gamma$   & $n^{*}$ & $\Gamma$ \\  
  \hline
  \hline
  \rule{0pt}{3ex}$2s 2p 3s$  &  2.146     &  5.94[-3]    & 2.131  & 5.58[-3]  & 2.140  & 5.73[-3] \\
  $2s 2p 4s$  &  3.120     &  1.79[-3]    & 3.157  & 1.63[-3]  & 3.157  & 1.59[-3] \\
  $2s 2p 5s$  &  4.170     &  6.98[-4]    & 4.164  & 6.96[-4]  & 4.161  & 6.69[-4] \\
  $2s 2p 6s$  &  5.172     &  3.59[-4]    & 5.166  & 3.60[-4]  & 5.162  & 3.44[-4] \\
  $2s 2p 7s$  &  6.173     &  2.10[-4]    & 6.167  & 2.10[-4]  & 6.163  & 2.00[-4] \\
\end{tabular}
\end{ruledtabular}
\end{table}
%%%
\begin{table}[hbtp!]
\caption{Comparison of effective quantum number, $n^{*}$ and resonance width $\Gamma$ (a.u.), of $^2{S^e}$ resonances between $2s^2$ and $2s2p(^3{P^o})$ thresholds.}
\label{tab:boronrespar2Se}
\begin{ruledtabular}
\begin{tabular}{lcccccc}
               & \multicolumn{2}{c}{\ASTRA{}} & \multicolumn{2}{c}{\NEWSTOCK{}} & \multicolumn{2}{c}{\TAEC{}~\cite{Argenti2016}} \\
        Conf.  & $n^{*}$   & $\Gamma$      & $n^{*}$    & $\Gamma$   & $n^{*}$ & $\Gamma$ \\
  \hline
  \hline
  \rule{0pt}{3ex}$2s 2p 3p$  &  2.765    &  4.99[-4]   & 2.743   &  5.22[-4]  & 2.733  & 4.70[-4]  \\
  $2s 2p 4p$  &  3.767    &  1.81[-4]   & 3.754   &  1.89[-4]  & 3.736  & 1.67[-4]  \\
  $2s 2p 5p$  &  4.766    &  8.51[-5]   & 4.756   &  8.96[-5]  & 4.736  & 7.75[-5]  \\
  $2s 2p 6p$  &  5.765    &  4.67[-5]   & 5.756   &  4.93[-5]  & 5.736  & 4.21[-5]  \\
  $2s 2p 7p$  &  6.765    &  3.15[-5]   & 6.756   &  2.99[-5]  & 6.735  & 2.54[-5]  \\
\end{tabular}
\end{ruledtabular}
\end{table}
%%%
\begin{table}[hbtp!]
\caption{Comparison of effective quantum number, $n^{*}$ and resonance width $\Gamma$ (a.u.), of $^2{D^e}$ resonances between $2s^2$ and $2s2p(^3{P^o})$ thresholds.}
\label{tab:boronrespar2De}
\begin{ruledtabular}
\begin{tabular}{lcccccc}
               & \multicolumn{2}{c}{\ASTRA{}} & \multicolumn{2}{c}{\NEWSTOCK{}} & \multicolumn{2}{c}{\TAEC{}~\cite{Argenti2016}} \\
        Conf.  & $n^{*}$   & $\Gamma$      & $n^{*}$    & $\Gamma$   & $n^{*}$ & $\Gamma$ \\
  \hline
  \hline
  \rule{0pt}{3ex}$2s 2p 3p$  &  2.581    &  1.95[-3]   & 2.579   &  1.88[-3]  & 2.568  & 1.74[-3]  \\
  $2s 2p 4p$  &  3.596    &  6.75[-4]   & 3.594   &  6.46[-4]  & 3.585  & 5.68[-4]  \\
  $2s 2p 5p$  &  4.601    &  3.12[-4]   & 4.599   &  2.98[-4]  & 4.585  & 2.87[-4]  \\
  $2s 2p 6p$  &  5.601    &  1.70[-4]   & 5.601   &  1.62[-4]  & 5.588  & 1.54[-4]  \\
  $2s 2p 7p$  &  6.603    &  1.06[-4]   & 6.601   &  9.77[-5]  & 6.588  & 9.34[-5]  \\
  \\
  $2s2p4f $   &   4.013   & 1.56[-7]  &  4.013  &  1.07[-7]  & 4.009  & 1.57[-7]  \\
  $2s2p5f $   &   5.012   & 1.37[-7]  &  5.012  &  9.73[-8]  & 5.009  & 1.19[-7]  \\
  $2s2p6f $   &   6.012   & 1.32[-7]  &  6.011  &  7.27[-8]  & 6.008  & 8.65[-8]  \\
\end{tabular}
\end{ruledtabular}
\end{table}

In \ASTRA{}, the resonance parameters are obtained by diagonalizing the multi-channel CC Hamiltonian in the presence of a CAP, whereas \NEWSTOCK{} enforces outgoing boundary conditions using exterior-complex scaling (ECS)~\cite{Simon1979,Lindroth1995,McCurdyJPB2004}.  
Table~\ref{tab:boronrespar2Po} compares the effective quantum number, $n^*=[2(E_{thr}-E_n)]^{-1/2}$, and resonance width, $\Gamma$, of selected autoionizing states with $^2P^o$ symmetry, between the the $2s^2 (^1{S^e})$ and $2s2p (^3{P^o})$ B$^+$ thresholds. 
The results for the dominant $2s2p (^3{P^o})ns$ series 
from \ASTRA{} are in excellent agreement with those obtained with \NEWSTOCK{} and \TAEC{}. 
Table~\ref{tab:boronrespar2Se} shows the parameters for the first few terms of the $2s2p (^3{P^o})np\,({^2S^e})$ autoionizing states. The resonance parameters for the two main series, $2s2p (^3{P^o})np$ and $2s2p (^3{P^o})nf$ of $^2{D^e}$ autoionizing states are listed in Table~\ref{tab:boronrespar2De}. 

\begin{figure}[htb]
\includegraphics[width=\linewidth]{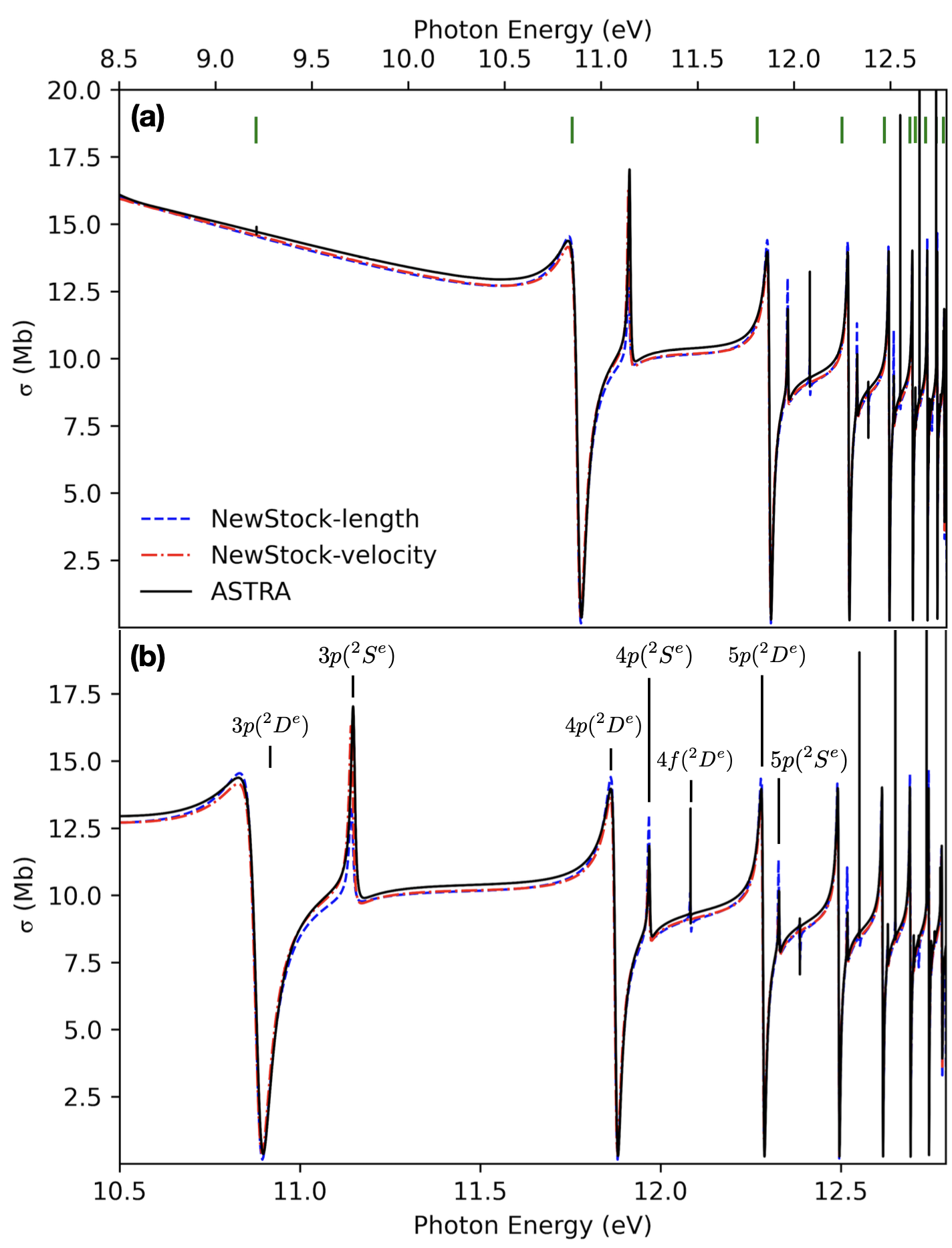}
\caption{(Color online)
(a) Total photoionization cross-section between the $2s^2 (^1S^e)$ and $2s2p({^3P^o})$ thresholds, from ${^2P^o}$ ground state of atomic boron, computed with \ASTRA{} (solid black line) and compared with the length (dashed blue line) and the velocity (dash-dotted red line) gauge results from \NEWSTOCK{}. The solid green lines indicate the positions of the bound states of $^2P^e$ symmetry. (b) Same cross-section as in the top panel, magnified in the resonant region, which exhibits autoionizing states with both $^2S^e$ and $^2D^e$ symmetry. The notation $n\ell(^2L^e)$ denotes a resonance with dominant configuration $2s 2p n\ell$ coupled to $^2L^e$ symmetry. \label{Fig:CrossSectionBoron}
}
\end{figure}
Figure~\ref{Fig:CrossSectionBoron} compares the total photoionization cross-section of boron, from the $^2{P^o}$ ground state to the energy region between the $2s^2$ and the $2s2p$ B$^+$ thresholds, computed in the length gauge with \ASTRA{}, with the one computed with \NEWSTOCK{}, both in length and in velocity gauge. As discussed in Sec.~\ref{sec:Method}, in this work, \ASTRA{} evaluates the total cross section using the optical theorem~\eqref{eq:opticaltheoremCAP}, whereas the cross section in \NEWSTOCK{} is computed from the dipole transition matrix elements  between the ground state and a complete set of scattering states of the atom. The two calculations, which differ entirely in their methodology, are in remarkable agreement with each other. In particular, \ASTRA{} accurately reproduces the characteristic asymmetry of the multi-channel resonant profiles of the many autoionizing states in this region. 

Figure~\ref{Fig:CrossSectionBoron}a covers the whole energy interval between the first and second threshold.  Figure~\ref{Fig:CrossSectionBoron}b shows a close up on the resonant region, and identifies selected terms in the three main resonant series. The broad $2s2pnp$\,$({^2D^e})$ and the narrow $2s2p3p (^2S^e)$ series dominate the total cross section. A much narrower $2s2pnf (^2D^e)$ series is visible near the second ionization threshold. 
As shown in Fig.~\ref{Fig:CrossSectionBoron}a, the boron spectrum computed with \ASTRA{} exhibits a few additional ultra-narrow resonant features, highlighted by vertical tics (green online), that deserve some comments. By comparing the positions of the Rydberg bound states in \NEWSTOCK{} and \ASTRA{}, it can be shown that this series corresponds to $^2P^e$ bound states. These bound states manifest themselves as vanishingly narrow resonances in the photoionization spectrum due to the miniscule coupling of the $^2P^e$ bound states to the $^2D^e$ continuum. Similar minor symmetry mixing are to be expected in a molecular code because, in $D_{2h}$ symmetry, the distinction between some $SO(3)$ irreducible representations is enforced dynamically, through numerical cancellations, rather than geometrically. That said, the mixing is small indeed. The largest of these resonances, at $\sim9.2$ eV, has a width of just $\approx 10^{-13}$~a.u., comparable to machine precision, whith the higher terms of the series being narrower still. Far from representing a failure of the code, therefore, such small values testify the remarkable accuracy with which \ASTRA{} does  reproduce the spherical symmetry of the atom.

The agreement between the two different \textit{ab-initio} approaches shows that the \ASTRA{} formalism, based on non-standard high-order TDMs between ionic states with different multiplicities and designed for molecular systems, is consistent and can be used to obtain accurate results for atoms~\cite{Puskar2023} as well.

\subsection{Nitrogen Molecule}

The nitrogen molecule is an ideal system to test the accuracy of states in the ionization continuum, thanks to the availability of several theoretical and experimental benchmarks in the literature. In this subsection, we report on several comparisons, within the fixed-nuclei approximation (FNA). First, we ascertain that, to the minimal CIS level, \ASTRA{} exactly reproduces the same result as an independent \emph{ad hoc} CIS code, as it should. Next, we test \ASTRA{} consistency and performance with the CC space constructed from correlated ions by comparing our results with those obtained with \XCHEM{}~\cite{MaranteJCTC2017}, a state-of-the art molecular-ionization code, as well as with experimental values.

\subsubsection{Comparison with \emph{ad hoc} CIS Code}
The \ASTRA{} suite has been benchmarked against an \emph{ad hoc} CIS code for the N$_2$ molecule, described in App.~\ref{app:CIS}. 
In the CIS model, the ionic states for the CC expansion are obtained by creating a vacancy in any of the valence orbitals of the Hartree-Fock determinant of the N$_2$ ground state. All these single-configuration ions are subsequently coupled to all available VMOs and VHOs. The CC space obtained with this approach, therefore, coincides with the CIS space from the HF ground state of N$_2$. 
The HF ground state was computed using a 6-31G basis, whereas for the hybrid orbitals we used B-splines of order 6, with a node separation of 0.77 a.u. and angular momentum $\ell\leq2$, within a 20 a.u. quantization box. 
The formulas for the transition-density matrices in the CIS basis, and the CC matrix elements of the Hamiltonian and the dipole operators, are particularly simple, as shown in App.~\ref{app:CIS}. The density matrices were used to test, in the CIS case, the consistency of the TDMs generated by \LUCIA{}, as well as that of the spin-adapted reduced TDMs evaluated by \ASTRA{}. 
The matrix elements of the Hamiltonian, computed separately by the CIS code, were used to ascertain the correctness of the \ASTRA{} structure code. The singlet states energy obtained from the Hamiltonian diagonalization in \ASTRA{} agree with those computed with the CIS code within $10^{-12}$~a.u., which is compatible with the propagation of error entailed by the necessary operations, executed with machine precision.

\subsubsection{Comparison with \XCHEM{}  in CI-singles subspace}

\ASTRA{} accounts for the matrix elements between poly-centric GTOs and B-splines, as well as for the exchange terms between hybrid and ionic orbitals. This feature allows us to use B-splines across the whole radial range, thus providing the flexibility necessary to achieve the large photoelectron energies typical of core spectroscopies (several hundreds eV) as well as to describe the continuum of comparatively large molecules. 

As detailed in the previous section, we ascertained that \ASTRA{} predictions coincide with those of an \emph{ad-hoc} CIS code, when the same hybrid integrals are used. 
In the present section, we show that, in the CIS case, the results generated by \ASTRA{} are compatible with those computed with \XCHEM{}, an independent and well established suite of molecular photoionization codes. 
Like \ASTRA{}, \XCHEM{} is based on the CC \textit{ansatz} and on the use of Gaussian-B-spline hybrid functions~\cite{MarantePRA2017,MaranteJCTC2017,KlinkerPRA2018,KlinkerJPCL2018,Borras2021}. In contrast with \ASTRA{}, however, \XCHEM{} employs a different algorithm to evaluate CC matrix elements, and it confines B-splines beyond a certain distance $R_0$ to limit their overlap with the ionic poly-centric GTOs. In \XCHEM{}, the molecular and the external regions are bridged by a diffuse set of monocentric GTOs $G_{i\ell k}(r)=r^{\ell+2k}\exp{(-\alpha_i r^2)}$~\cite{Marante2014}. 

Table~\ref{tab:ASTRA_dipoles} compares the energy of the N$_2$ ground state and of the first few excited singlet states, as well as the dipole oscillator strengths from the former to the latter, obtained by diagonalizing the CC Hamiltonian computed either with \XCHEM{}, using the 6-31G basis, or with \ASTRA{}, using the 6-31G and the larger cc-pVQZ basis. In all these cases, the CC expansion includes the $X\phantom{A}\!\!\!^2\Sigma_g^+$, $A\phantom{A}\!\!\!^2\Pi_u$ and $B\phantom{A}\!\!\!^2\Sigma_u^+$ parent ions, augmented by single-electron states with $\ell\leq 2$, in a 400 a.u. quantization box. The ions were confined to a sphere with radius $R_\mathrm{mol}=20$ a.u. and two CAP were employed, starting at 100 and 200 a.u. with a strength of $10^{-6}$ and $10^{-5}$, respectively. The ionic states are obtained from the HF configuration by removing one electron from the $3\sigma_g$, $1\pi_u$, and $2\sigma_u$ orbitals. The B-splines were of order 8, the node separation chosen as $\Delta r=0.5$ a.u. and for the diffuse GTOs in \XCHEM{} we followed the same even-tempered set of exponents $\alpha_i$ defined in~\cite{KlinkerJPCL2018}, with $k=2$. The nitrogen molecule is aligned along the $\hat{z}$ axis and the internuclear distance is set to the experimental equilibrium value $R=1.098$\,Å~\cite{Huber1979}.

\begin {table}[hbtp!]
\normalsize
\caption{\label{tab:ASTRA_dipoles}
Energy of and dipole oscillator strength (OS) between the ground state and the first few bright states of $\mathrm{N_2}$, computed with \XCHEM{} and \ASTRA{} in the CI-singles limit, using a 6-31G basis for the HF orbitals. The energies are relative to \ASTRA{} ground state.}
\begin{ruledtabular}
\begin{tabular}{rcccc}
 & \multicolumn{2}{c}{Energy (eV)} &  \multicolumn{2}{c}{ OS (a.u.) }  \\ \cline{2-3}\cline{4-5}
\rule{0pt}{3ex}State& \XCHEM{} &  \ASTRA{} &  \XCHEM{} &  \ASTRA{} \\  
\tableline
\rule{0pt}{3ex} $X\,{^1\mathrm{\Sigma_{g}^+}}$  & \phantom{0}0.004 &	\phantom{0}0.000 &- & -  \\
$1\,\,{^1\mathrm{\Pi_{u}^{\phantom{+}}}}$  & 15.236	& 15.236  & 0.2062 & 0.2066  \\
$1\,\,{^1\mathrm{\Sigma_{u}^+}}$  & 16.267 &	16.287  & 0.2053 & 0.2098  \\
$2\,\,{^1\mathrm{\Pi_{u}^{\phantom{+}}}}$   & 16.311 & 16.321  & 0.0812 & 0.0802  \\
\end{tabular}
\end{ruledtabular}
\end{table}
The energies computed with \XCHEM{} and \ASTRA{} using the 6-31G basis differ by no more than 0.02~eV, which is a remarkable agreement, given how different the virtual monoelectronic orbitals in the molecular region are for these two methods. The oscillator strength is more sensitive to the quality of the underlying electronic numerical basis. Nevertheless, \ASTRA{}'s results do not deviate more than 1\% from those of \XCHEM{}. 

\begin{figure}[hbtp!]
\includegraphics[width=\linewidth]{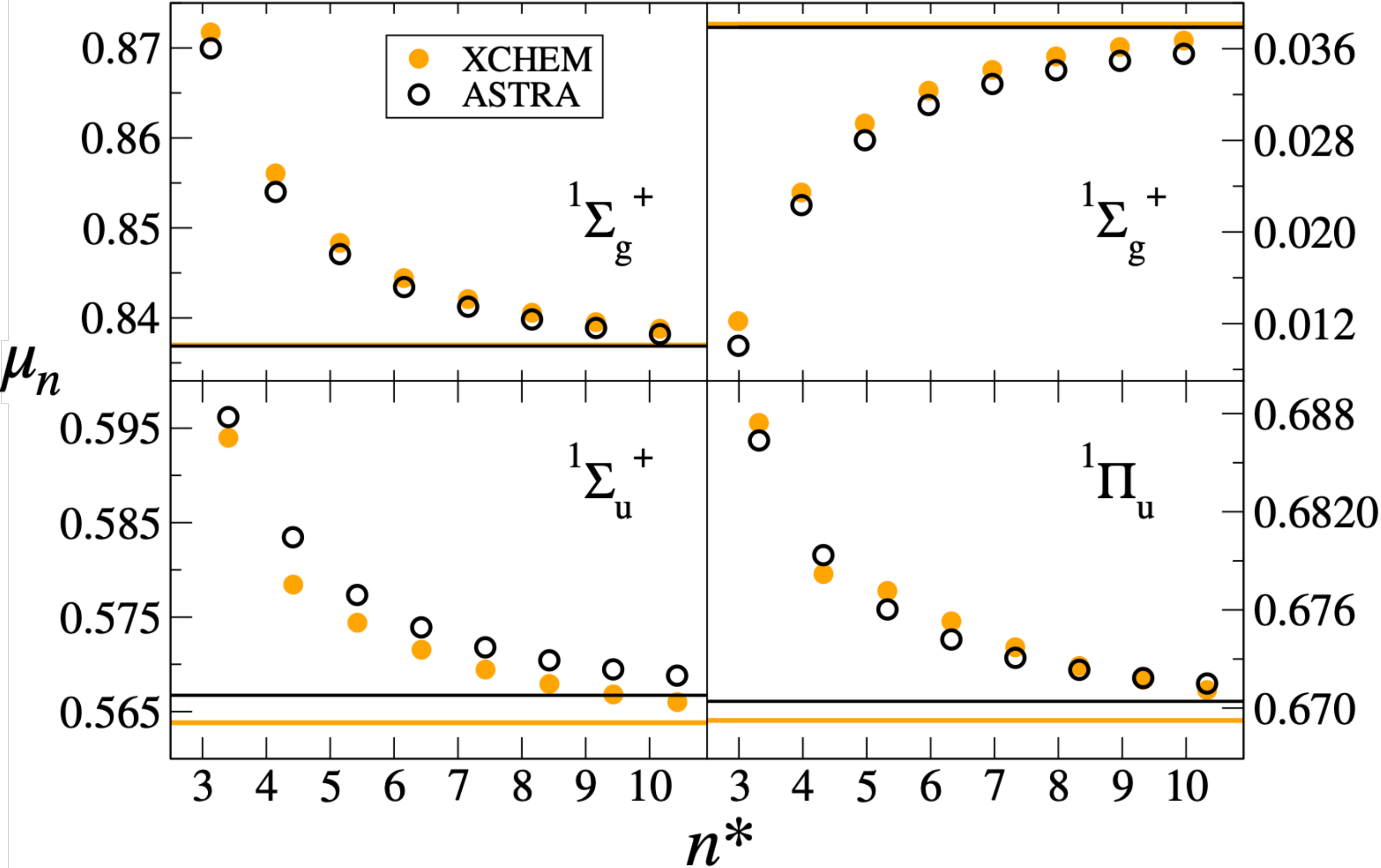}
\caption{\label{Fig:QD}
Quantum defect $\mu_n^\Gamma$ \textit{vs} the effective principal quantum number $n^*$ for four Rydberg series of bound states in three symmetries, computed with \XCHEM{} (filled circles, yellow online) and \ASTRA{} (black empty circles) within the same CI-singles CC space used in Table~\ref{tab:ASTRA_dipoles}. Horizontal lines indicate $\mu_\infty^\Gamma = \lim_{n\to\infty}\mu_n^\Gamma$.
}
\end{figure}

In Fig.~\ref{Fig:QD}, we compare the quantum defect $\mu_n\equiv n-n^*$ of four Rydberg series converging to the first ionization threshold, $X\,{^2\Sigma_g^+}$.
The quantum defects obtained with the two methods are within $0.5\%$ for all values of $n^*$. 
The agreement, therefore, is not restricted to the first few bound states, which may be well represented by GTOs alone, but it extends to Rydberg states with large quantum numbers and with a radial size well beyond the molecular region.
These positive comparisons shows that, for the present calculation: i) the two different configuration spaces made available to the $N$-th electron in \XCHEM{} and \ASTRA{} are equivalent, and ii) \ASTRA{} correctly employs the \GBTOlib{} hybrid basic integrals.

\subsubsection{CC with correlated ions}\label{subsec:N2_Correlated_Ions}

In this section, we go beyond the CIS approximation by considering the CC expansion obtained from the $X\,{\Sigma_g^+}$, $A\,{^2\Pi_u}$ and $B\,{^2\Sigma_u^+}$ correlated ions augmented by virtual orbitals with asymptotic angular momentum up to $\ell_{max}=3$. The ions are computed by means of a CASCI calculation~\cite{MEST2000}, over the $\{2\sigma_{g/u},3\sigma_{g/u},1\pi_u,1\pi_g\}$ set of active orbitals, keeping the core $1\sigma_{g/u}$ orbitals inactive (doubly occupied), and employing three different GTO bases: 6-31G, cc-pVDZ, and cc-pVTZ. Table~\ref{tab:ASTRA_bounds} compares the energies of the first five singlet bound states obtained with \ASTRA{} in the three different bases, alongside those obtained with \XCHEM{} in the minimal 6-31G basis. All the energies are relative to the ground state obtained with \ASTRA{} with the cc-pVTZ basis. For the \XCHEM{} calculations, the same active space defined for \ASTRA{} is employed to perform a CASCI and obtain the correlated ionic states. The same basis of diffused GTOs utilized in the CIS calculations is used to complement the polycentric GTOs at short range. Once again, the agreement between \XCHEM{} and \ASTRA{} is excellent. 

\begin {table}[hbtp!]
\normalsize
\caption{\label{tab:ASTRA_bounds}
Energy of the first five singlet states of $\mathrm{N_2}$, in eV, obtained by including correlated parent ions in the CC (see text for details) and using different GTO basis. The energy values are with respect to the \ASTRA{} ground state when the basis set used to generate the molecular orbitals is cc-pVTZ.}
\begin{ruledtabular}
\begin{tabular}{rccccc}
 & \XCHEM{} & \multicolumn{3}{c}{\ASTRA{}}  \\ \cline{2-2}\cline{3-5}
\rule{0pt}{3ex}\textbf{State} &  
\multicolumn{1}{c}{\textbf{6-31G}} & 
\multicolumn{1}{c}{\textbf{6-31G}} & \multicolumn{1}{c}{\textbf{cc-pVDZ}} &  \multicolumn{1}{c}{\textbf{cc-pVTZ}} \\
\tableline
\rule{0pt}{3ex}
$1\,\,{^1\mathrm{\Sigma_{g}^+}}$   & 0.639 & 0.631 & 0.128 & 0.000 \\
$1\,\,{^1\mathrm{\Pi_{g}^{\phantom{+}}}}$      & 12.890	& 12.889 & 10.841 & 10.219  \\
$1\,\,{^1\mathrm{\Sigma_{u}^-}}$& 13.142 & 13.138 &	11.078 & 10.467  \\
$2\,\,{^1\mathrm{\Sigma_{g}^+}}$   & 15.051 & 15.051 & 13.038 & 12.784 \\
$1\,\,{^1\mathrm{\Delta_{u}^{\phantom{+}}}}$   & 15.636 & 15.639 & 13.622 & 13.411 \\
\end{tabular}
\end{ruledtabular}
\end{table}

\begin{figure}[tb]
\includegraphics[width=\linewidth]{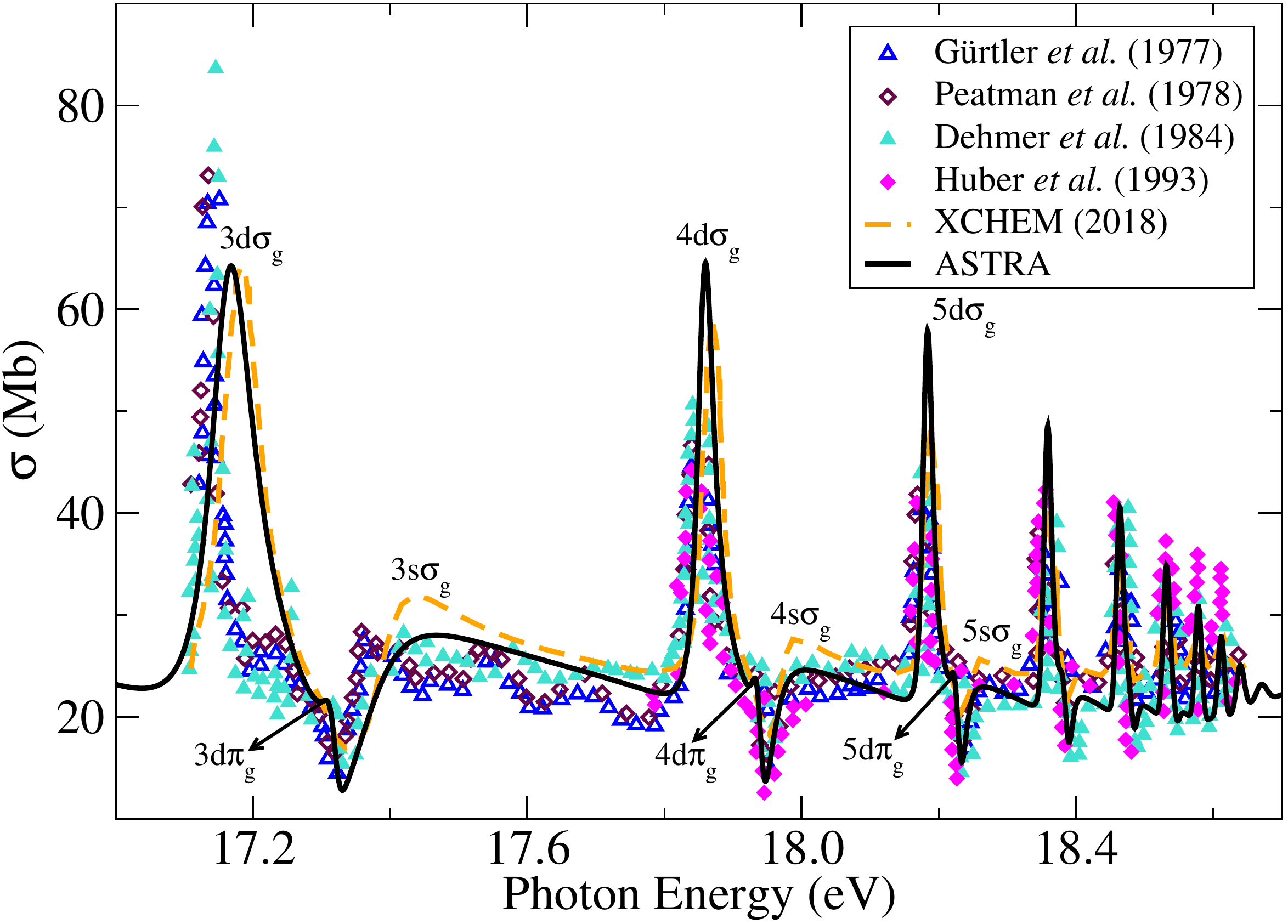}
\caption{(Color online)
Total photoionization cross section of N$_2$ in the Hopfield resonances region. The spectrum computed with \ASTRA{} (solid black line) is compared with four independent measurements (triangles and diamonds)~\cite{Gurtler1977,Peatman1978,Dehmer1984,Huber1993} and another theoretical spectrum obtained with \XCHEM{} (dashed line)~\cite{KlinkerJPCL2018}. \label{Fig:N2_CrossSec}
}
\end{figure}
In the present implementation, the CC expansion only consists of correlated ionic states augmented with an extra electron either in a bound or unbound orbital. A truncated CC expansion may be insufficient to accurately reproduce all the resonant states in the energy region of interest. This is because, in the CC expansion, the response of the ionic bound electrons to an additional electron is described only in terms of the expansion on a small set of states. While the CC captures well the polarization of the ion when the additional electron is located at several Bohr radii from the nucleus, it cannot describe well dynamic correlation at short-range, such as the Coulomb hole~\cite{Sule1995}, which has components in the multiple-ionization states of the ion itself.  
A standard way of dealing with this deficiency of the CC approach is to extend the expansion to include a pseudo-channel with a large number of $N$-electron configurations built out of localized orbitals, without constraining $N-1$ electrons to any specific set of eigenstates of the ionic Hamiltonian. This pseudo channel, which is sometimes referred to as the $\mathcal{Q}$ space~\cite{Feshbach1958}, can be very large. The \CKMESA{} code~\cite{Douguet2018a} has an efficient way of including the $\mathcal{Q}$ space via an optical potential built entirely out of GTOs~\cite{MESA_code}. \XCHEM{} can also account for the $\mathcal{Q}$ space, even if it does so explicitly, which is more demanding than via an optical potential~\cite{Douguet2018a}. The inclusion of the $\mathcal{Q}$ space in \ASTRA{} is outside the scope of the present work. One viable approach to circumvent this limitation is extending the CC expansion so to include closed ionization channels whose parent ions are augmented only by AMO, VMO, and VHO, and in which the extra parent ions do not have to accurately reproduce any eigenstates of the ion. 

We followed this approach to reproduce the high-resolution synchrotron radiation photoionization spectrum of N$_2$, in the vicinity of the Hopfield series of autoionizing states~\cite{KlinkerJPCL2018}. Besides the three parent ions $X\phantom{A}\!\!\!^2\Sigma_g^+$, $A\phantom{A}\!\!\!^2\Pi_u$ and $B\phantom{A}\!\!\!^2\Sigma_u^+$ we have used in the previous calculations, we added another 15 to the CC: $2\phantom{A}\!\!\!^2\Sigma_{g}^+$, $1\phantom{A}\!\!\!^2\Sigma_{g}^-$, $\varA{1-2}\phantom{A}\!\!\!^2\Delta_{g}$, $1\phantom{A}\!\!\!^2\Phi_{g}$, $\varA{2-4}\phantom{A}\!\!\!^2\Pi_{u}$, $2\phantom{A}\!\!\!^2\Sigma_{u}^+$, $\varA{1-2}\phantom{A}\!\!\!^2\Delta_{u}$, $\varA{1-3}\phantom{A}\!\!\!^2\Pi_{g}$ and $1\phantom{A}\!\!\!^2\Sigma_{u}^-$, to define the close ionization channels. In Fig.~\ref{Fig:N2_CrossSec} we show the comparison of the photoionization cross section computed with \ASTRA{}, the \XCHEM{} result reported in~\cite{KlinkerJPCL2018} alongside four experimental spectra~\cite{Gurtler1977,Peatman1978,Dehmer1984,Huber1993}. The theoretical signals have been convoluted with a 0.015 eV width normalized Gaussian to account for the experimental resolution. In the photoionization spectrum we can distinguish the three Hopfield series of resonances converging to the $B\phantom{A}\!\!\!^2\Sigma_u^+$ ionization threshold: $ns\sigma_g$, $nd\sigma_g$ and $nd\pi_g$, for which we indicate the first three terms. The \ASTRA{} result is in very good agreement with both the \XCHEM{} calculation and with the experimental spectra. For the first resonant feature corresponding to $3d\sigma_g$, both theoretical the results show a broader peak than the measured value, which is likely due to residual electronic correlation missing in the early terms of the Rydberg series. 

The size of the close-coupling space, in this case, is 16418 for the $A_g$ $D_{2h}$ symmetry, and 10918 for all the other $D_{2h}$ symmetries considered. The calculations were conducted on a standalone workstation equipped with two 2.8GHz Intel Xeon Platinum-8362 processors (32 cores each). While the current implementation of \ASTRA{} is serial, \textsc{lapack}~\cite{lapack99} diagonalization routines employ multithreaded drivers that occupied, on average, between 17 and 19 cores. The calculation of the CC matrix elements for all the operators, the full diagonalization of the real and complex Hamiltonian and the calculation of the photoionization cross section took  $\sim$2 hours. 

\subsection{Formaldehyde}
To test \ASTRA{} on a non-linear polyatomic molecule, we selected formaldehyde, H$_2$CO, for which resonably accurate experimental~\cite{Cooper1996,Tanaka2017} and theoretical~\cite{Cacelli2001} photoionization cross sections are available. As for N$_2$, we used the FNA at the equilibrium geometry: $R_{\mathrm{CO}}=1.21$\,\AA, $R_{\mathrm{CH}}=1.12$\,\AA, $\measuredangle \mathrm{HCH}=116.5^\mathrm{o}$~\cite{Wang2022,Gurvich1989}. 
As detailed below, we conducted calculations with \ASTRA{} at several levels of accuracy. In our largest calculation, the molecular orbitals are optimized by a state average MCSCF calculation on the first ten neutral state of the molecule, the VHO and ESO orbitals are computed with $\ell\leq 4$, and the CC expansion includes the first 37 molecular ions. The first ten ions are augmented with the full set of internal and external orbitals (AMO, VMO, VHO, and ESO), whereas the remaining 27 ions, which are all closed in the energy region of interest, are augmented by the internal orbitals only (AMO, VMO, and VHO).

\begin{figure}[hbtp!]
\includegraphics[width=\linewidth]{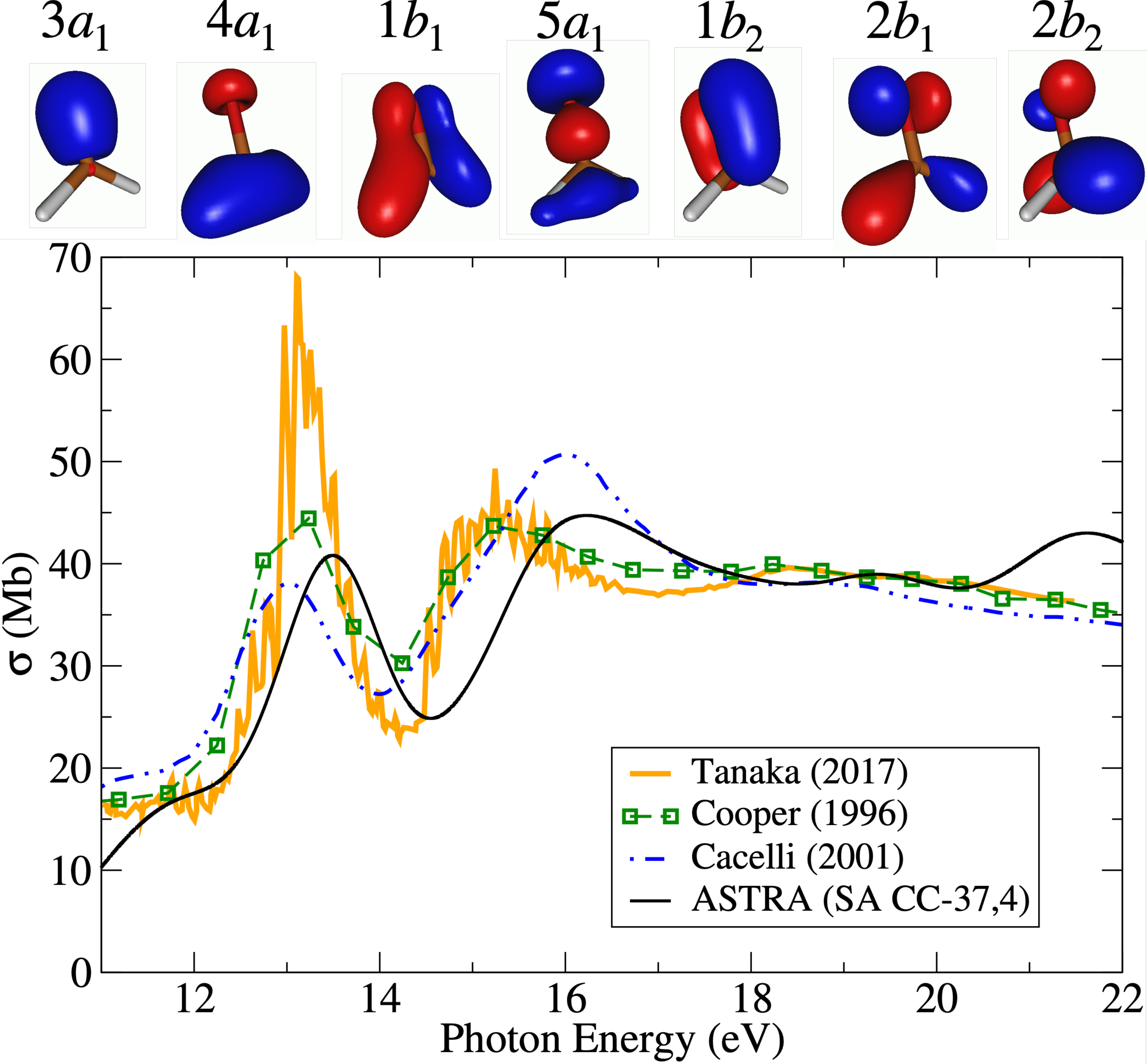}
\caption{ (Color online)
Formaldehyde total photoionization cross section above the ionization threshold. The \ASTRA{} result (black-solid line) is obtained by including 37 ions in the CC expansion with $\ell\leq 4$. The ions were obtained through a CASCI using the orbitals coming from a SA-MCSCF calculation (see text for details). The \ASTRA{} sepctrum is compared with the RPA calculation in~\cite{Cacelli2001} (dotted-dash line) and two measurements, with lower (dashed line and squares) and higher spectral resolution (orange-solid line)~\cite{Cooper1996,Tanaka2017}. The top of the figure shows the first seven molecular orbitals included in the \ASTRA{} active space, in ascending energy order, up to the LUMO ($2b_2$)~\cite{MOLDEN2000,MOLDEN2017}. \label{Fig:Formaldehyde_CrossSec}  
}
\end{figure}

Figure~\ref{Fig:Formaldehyde_CrossSec} compares the photoionization cross section of formaldehyde computed with \ASTRA{} between 11~eV and 22~eV, using the highest level of electronic correlation described above, with Cacelli~\emph{et al.} RPA calculation~\cite{Cacelli2001}, and with two measurements: one from a lower-resolution 1996 experiment by Cooper~\emph{et al.}~\cite{Cooper1996} and the other from a higher-resolution 2017 experiment by Tanaka~\emph{et al.}~\cite{Tanaka2017}. 
The spectrum exhibits two prominent shape resonances, a narrower one near 13~eV, and a broader less pronounced one at 15~eV. In the higher-resolution experiment by Tanaka, the first peak has smaller width and larger height. Considering the different resolutions, however, the results of the two experiments are compatible with each other.
The results from \ASTRA{} are convoluted with a 1.2~eV Gaussian window, the same as in~\cite{Cacelli2001}, in line with the $\approx5\%$ resolution reported by Cooper~\cite{Cooper1996}.
The RPA calculation~\cite{Cacelli2001} predicts well the position of the first peak, which it underestimates by just $0.2$~eV, while it overestimates the position of the second peak by about 1~eV. Furthermore, the height of the first and second peak are significantly under- and over-estimated, respectively. These results suggest that both resonances are highly sensitive to correlation. 
In the \ASTRA{} results, the position of the first peak is overestimated by about 0.2~eV, but its magnitude is in better agreement with the experiment by Cooper. Similarly to Cacelli's result, the position of the second peak predicted by \ASTRA{} is about 1~eV above the experimental value. The peak height, however, is almost coincident with one observed in the experiment. 

The results obtained with \ASTRA{} are arguably the ones in best agreement with the experiment, to date. However, they also suggest that the theoretical model used here is still incomplete. Two possible causes for the discrepancy with the experiment are the residual dynamic correlation not captured by the CC expansion, and the fixed-nuclei approximation, which prevents us from accounting for the effects of nuclear motion and rearrangement in the molecular ion. On the electronic-structure side, ideally, the CC expansion should include a flexible and large pseudo-channel formed by arbitrary localized configurations. On the nuclear side, one should treat consistently nuclear motion and nuclear rearrangement on multiple excited electronic potential energy surfaces in a molecule with several degrees of freedom. While we plan to tackle both of these challenges in the future, these are formidable tasks beyond the scope of the present work.

\begin{figure}[hbtp!]
\includegraphics[width=\linewidth]{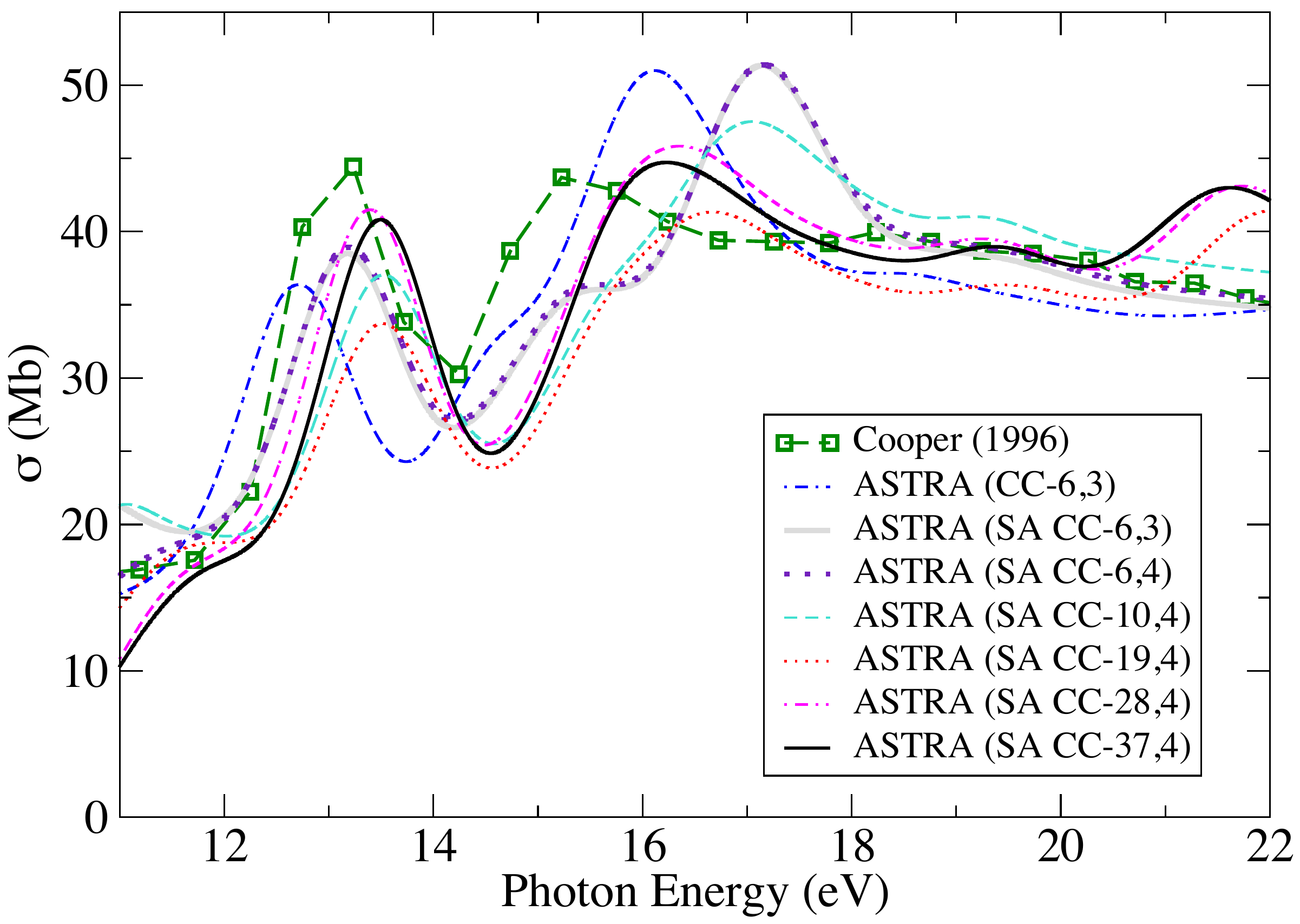}
\caption{ (Color online)
Formaldehyde total photoionization cross section above the ionization threshold. Seven \ASTRA{} spectra are shown together with Cooper's experiment ($\Delta E\approx 1.2$ eV)~\cite{Cooper1996}. Six of the \ASTRA{} calculations employed the same SA-MCSCF orbitals to perform the CASCI for the ionic states and they differ in the number of ions included in the CC expansion and in the maximum angular momentum considered for the partial waves. The format SA CC-N,L means that the calculation was performed by using the SA orbitals and a CC expansion with N ions and $\ell\leq L$. \label{Fig:Formaldehyde_CrossSec_Convergence}  
}
\end{figure}

To illustrate the effect of the CC basis on formaldehyde photoionization cross-section, within the current capabilities of \ASTRA{}, in Fig.~\ref{Fig:Formaldehyde_CrossSec_Convergence} we compare the results obtained with seven different calculations, carried out with progressively more demanding parameters.
In the first calculation, the CC expansion comprises six ionic states, $\varA{1-2}\!\!\!\phantom{A}^2B_1$, $1\!\!\!\phantom{A}^2B_2$, $\varA{1-3}\!\!\!\phantom{A}^2A_1$, obtained from a CASCI calculation of 12 electrons in the 9 active orbitals, $\{3a_1,4a_1,1b_1,5a_1,1b_2,2b_1,2b_2,6a_1,3b_1\}$, whereas the two core orbitals $1a_1$ and $2a_1$ (corresponding to the O$1s$ and C$1s$ K shells) were doubly occupied. These orbitals are obtained from the HF calculation of the neutral molecule within the cc-PVTZ GTO basis.
To build the CC channels, each of these ions was augmented by an electron in any of the GTO (AMO, in the notation of Sec.~\ref{sec:Method}) and hybrid orbitals (VHO), and B-splines with asymptotic angular momentum $\ell\leq3$ spanning the whole quantization box (ESO). The B-spline and $V_{\textsc{CAP}}$ parameters coincide with those used for nitrogen. 

This first \ASTRA{} calculation reproduces the characteristic double-hump structure of the spectrum and it is in very good agreement with the results by Cacelli~\emph{et al.}~\cite{Cacelli2001}. In particular, the absolute magnitude of the two peaks in the two independent calculations are remarkably similar. Both theoretical calculations, however, underestimate the position and magnitude of the first peak, and overestimate the energy and magnitude of the second peak, compared with the experiment.

In the other six \ASTRA{} calculations, the molecular GTO orbitals are optimized through a SA-MCSCF step over the first 10 neutral states, with equal weights. The ions are subsequently determined by a CASCI calculation with the same number and type of inactive and active orbitals as in the first case. 

In the second calculation, the CC space is generated from the first six ionic states only, augmented with orbitals with $\ell\leq 3$. The first and the second calculation, therefore, differ only in the optimization of the ion orbitals. The calculation reproduces very well the position of the first peak, quite possibly a coincidence, whereas the second peak splits into a shoulder around the position of the peak observed in the experiment and in a prominent third peak around 17~eV, with no correspondence in the measured data.

In the third calculation, the ions in the CC expansion coincide with those in the second calculation, but all the ions are now augmented with orbitals with $\ell\leq 4$. The second and the third calculation, therefore, differ only in the photoelectron partial-wave expansion. Both photoionization cross sections are almost identical except in the energy region approaching the ionization threshold ($E<12$~eV), which indicates that in the region of interest, the partial wave expansion is close to convergence for $\ell\leq 4$. 

The fourth calculation coincides with the third, except that the CC expansion is extended to include the first 10 ions, each augmented by both the internal (AMO, VMO, and VHO) and by the external (ESO) partial-wave orbitals. The shoulder observed in the two previous spectra is greatly reduced and the magnitude of the second peak decreases, being closer to the experiment. The position of the first peak gets shifted 0.2~eV to higher energies and it's magnitude decreases by 2~Mb.  

In the remaining three calculations, we investigated the convergence of the cross section using the same number of fully-augmented partial-wave channels as in the fourth, but adding 9, 18, and 27 other ions to the CC expansion, each augmented by the internal orbitals only. Since the corresponding channels are all closed in the energy region of interest, these last three calculations include progressively more dynamic correlation. Going from 10 to 19 to 28 ions, the position of the second peak is lowered by almost 1~eV. The position of the first peak is only marginally affected. The inclusion of nine other ions, from 28 to 37, has a comparatively small effect, which suggests that these calculations are close to convergence for the present choice of orbitals.

\section{Conclusions and Perspectives\label{sec:Conclusions}}

We have described a transition-density-matrix approach to the close-coupling method (TDMCC) for molecular photoionization, implemented in the \ASTRA{} code, and used it to compute the bound-state parameters, autoionizing-state parameters, and the single-photon integral photoionization cross section of benchmark atomic and molecular systems. The TDMCC method scales well with the size of the CI space of the parent ion, and it delivers results in excellent agreement with those in the literature.

In future works, we will explore the scalability of \ASTRA{} by studying the ionization process in increasingly larger systems. Preliminary calculations indicate that ASTRA is capable of simulating the ionization of molecules as large as metallo-porphyrines, organometallic compounds comprised of $37$ or more atoms, which opens a new way to the detailed predictions of wave-function-based methods for the ionization of molecules of biological interest.

From the methodological point of view, the \ASTRA{} program has several natural directions along which to evolve. First, is the calculation of multichannel scattering states, which will allow us to determine photoelectron angular distributions~\cite{Wu2011,Dowek2022}, either in stationary regime, or as a result of the interaction of a target molecule with an ultrashort pulse of radiation. Second, is the extension of the calculation of the ionic states and of their TDMs from the present CASCI level to the more general RASCI~\cite{Malmqvist1990,Lischka2018,Zhou2019,Casanova2021} level. This extension will be particularly useful to describe highly conjugated systems. 
Third, we plan to use the SACC space as the basis for the time-dependent description of the photoionization wavepackets generated by the interaction of a target molecule with a sequence of ultrashort ionizing pulses~\cite{Harkema2021,Leshchenko2023,Borras2021}. This direction will allow us to reproduce transient-absorption spectra~\cite{Kobayashi2019,KrausNatureReviews2018,Zinchenko2021,Rebholz2021}, four-wave mixing spectra~\cite{Cao2018,Warrick2018,Fidler2022}, and photoelectron spectra~\cite{Schuurman2022} for state-of-the-art experiments.

In the longer term, we plan to explore also the many other natural extensions of the TDMCC formalism of \ASTRA{}, such as the inclusion of a Q-space for the accurate description of short-range correlation~\cite{Schneider1991}, the inclusion of double-escape channels~\cite{Zaunerchyk2015,Mucke2015,Lablanquie2016}, to reproduce molecular double-ionization processes, as well as the description of multi-fragment systems~\cite{MurilloSanchez2018,Lee2021,Allum2022,Howard2023,Rolles2023} (e.g., a dissociating molecule, or a loosely bound aggregate) in terms of the TDMs of its separated components, in which the TDMs of the whole aggregate naturally factorizes, thus leading to drastic reduction in computational cost.

\section*{Acknowledgments}
This work is supported by the DOE CAREER grant No. DE-SC0020311. We are grateful to Dr.~Zdenek Masin for fruitful discussions and suggestions on the use of the \GBTOlib{} hybrid-integral library. We acknowledge generous allocation time at NERSC under the contract No. DE-AC02-05CH11231 and the award BES-ERCAP0024720, as well as time allocation from UCF Advanced Research Computing Center (ARCC). 
\appendix

\section{Second-quantization formalism}\label{app:SecondQuantization}

In this section we summarize the correspondence between antisymmetric electronic functions in the first-quantization coordinate representation and the associated Fock space, as well as some basic properties of many-body operators in second-quantization (SQ). For a general description of the SQ formalism, we refer the reader to the first chapter in \cite{MEST2000}.

Let's gather the spatial coordinate $\vec{r}$ and the spin coordinate $\zeta\in\left\{-\frac{1}{2},\frac{1}{2}\right\}$ of an electron in the coordinate $\mathrm{x}=(\vec{r},\zeta)$. The state space for a single electron is spanned by a basis of spin orbitals $\phi_P(\mathrm{x})=\varphi_p(\vec{r}){^2\chi_\pi(\zeta)}$, where $\varphi_p(\vec{r})\in\mathcal{L}^2(\mathbb{R}^3)$, $^2\chi_\pi(\zeta)=\langle \zeta|\pi\rangle= \delta_{\sigma\pi}$, and $P=(p,\pi)$ is a spin-orbital index, with $\pi\in\left\{-\frac{1}{2},\frac{1}{2}\right\}$. In an $N$-electron system, the state space is spanned by Slater determinants $|\mathbf{P}\rangle$ built from arbitrary selections $\mathbf{P}=(P_1,\ldots,P_N)$ of $N$ distinct spinorbitals from a complete one-particle basis, and defined as
\begin{eqnarray}
\langle \mathbf{x}|&&\mathbf{P}\rangle=\frac{1}{\sqrt{N!}}
\left|
\begin{array}{cccc}
\phi_{P_1}(\mathbf{x}_1)&\phi_{P_2}(\mathbf{x}_1)&\cdots&\phi_{P_N}(\mathbf{x}_1)\\
\phi_{P_1}(\mathbf{x}_2)&\phi_{P_2}(\mathbf{x}_2)&\cdots&\phi_{P_N}(\mathbf{x}_2)\\
\vdots&\vdots&\ddots&\vdots\\
\phi_{P_1}(\mathbf{x}_N)&\phi_{P_2}(\mathbf{x}_N)&\cdots&\phi_{P_N}(\mathbf{x}_N)
\end{array}
\right|\nonumber\\\nonumber\\
&&=\sqrt{N!}\, \hat{\mathcal{A}}\,\,
\left[
\phi_{P_1}(\mathbf{x}_1)\phi_{P_2}(\mathbf{x}_2)\cdots\phi_{P_N}(\mathbf{x}_N)
\right]=\nonumber\\
&&=\left|\phi_{P_1}\phi_{P_2}\cdots\phi_{P_N}\right|,
\end{eqnarray}
where $\mathbf{x}=(\mathrm{x}_1,\mathrm{x}_2,\ldots,\mathrm{x}_N)$ are the electronic coordinates, $\hat{\mathcal{A}} = \frac{1}{N!}\sum_{\mathcal{P}\in\mathcal{S}_N}(-1)^{p}\mathcal{P}$ is the antisymmetrizer, $\hat{\mathcal{A}}^2=\hat{\mathcal{A}}$, $\hat{\mathcal{A}}^\dagger=\hat{\mathcal{A}}$, $\mathcal{S}_N$ is the symmetric group (group of permutations of $N$ objects), and $\mathcal{P}$ and $p$ are a permutation and its parity, respectively. These antisymmetric functions can be expressed in an equivalent way with the SQ formalism.

In SQ, the state of a system is expanded on a basis of occupation states that specify, for each spinorbital in an ordered list, if it is occupied by an electron or not. A state in which spinorbitals 1, 3, 4, and 7 are occupied, for example, is indicated by the ket $|1,0,1,1,0,0,1,0,0,0,\ldots\rangle$. The total number of electrons in any such occupation state is given by $N_e=\sum_n n_i$, where $n_i$ are the occupation numbers. A state with no electrons is represented by the so-called vacuum state, $|-\rangle = |0,0,0,0,\ldots\rangle$, $\langle-|-\rangle=1$.
The space spanned by all occupation-number states with finite number of electrons is referred to as Fock space.
To describe the action of arbitrary observables in this space, it is convenient to introduce operators that rise or lower the occupancy of a given spinorbital. 
The annihilation operator $a_P$ removes an electron from the spinorbital $P$, if full, and annihilate the state otherwise. It acts on an occupation-number state as
\begin{equation}
a_P\,| k_1,\,\cdots, k_P, \cdots \rangle  =  \delta_{k_P\,1}\,\,\Gamma^{\mathbf{k}}_P\,\, | k_1,\,\cdots, \,0_P,\,\cdots\rangle,
\end{equation}
where $\Gamma^{\mathbf{k}}_P = (-1)^{k_1+k_2+\ldots+k_{P-1}}$. 
The adjoint of $a_P$, $a_P^\dagger$, adds an electron to spinorbial $P$, if empty, and annihilate the state otherwise,
\begin{equation}
a_P^\dagger | k_1,\,\cdots, k_P, \cdots \rangle  =  \delta_{k_P\,0}\,\,\Gamma^{\mathbf{k}}_P\,\, | k_1,\,\cdots, \,1_P,\,\cdots\rangle.
\end{equation}
Creation and annihilation operators satisfy the following well-known anti-commutation relations, 
\begin{equation}
\{a_I,a_J\}=\{a^\dagger_I,a^\dagger_J\}=0,\qquad \{a^\dagger_I,a_J\}=\delta_{IJ}.
\end{equation}
The correspondence between first and second-quantization formalism is completed by a phase convention for the equivalence between Slater determinants and occupation-number states. Here, we adopt the following convention
\begin{equation}
\left|\phi_{P_1}\phi_{P_2}\cdots\phi_{P_N}\right|\quad \leftrightarrow\quad 
a_{P_1}^\dagger a_{P_2}^\dagger \cdots a_{P_N}^\dagger | - \rangle.
\end{equation}

\subsubsection{Operators in second-quantization formalism}
In first quantization, the definition of one-body operators, $\langle \mathbf{x}|\hat{O}|\mathbf{x}'\rangle=\sum_{i=1}^{N_e} o(x_i;x_i')$, and of the two-body Coulomb repulsion operator, $\langle \mathbf{x},\mathbf{x}'|\hat{G}|\mathbf{x},\mathbf{x}'\rangle=\sum_{i,j<i} |\vec{r}_i-{\vec{r}_j}'|^{-1}$, depend explicitly on the number of particles in the system. In SQ, on the other hand, these operators do not explicitly depend on the number of electrons, which is a major advantage,
\begin{eqnarray}
\hat{O}&=&\sum_{RS}o_{RS} a^\dagger_R a_S,\\
\hat{G}&=&\frac{1}{2}\sum_{PQRS}[PQ|RS]\, a^\dagger_P a^\dagger_R a_S a_Q.
\end{eqnarray}
where
\begin{eqnarray}
o_{RS}&=&\iint dx\, dx'\, \phi^*_R(x) o(x,x') \phi_S(x'),\nonumber\\
\left[PQ|RS\right]&=&\iint dx\, dx'\,\frac{\phi_P^*(x)\phi_Q(x)\,\phi_R^*(x')\phi_S(x')}{r_{12}}.\nonumber
\end{eqnarray}

In \ASTRA{}, we need to evaluate matrix elements of strings of creation and annihilation operators between states $|A\rangle$ and $|B\rangle$ with well-defined spin, $S_A$, $S_B$, and spin projections, $\Sigma_A$, $\Sigma_B$, respectively. To this end, it is convenient to expand the operators in spherical tensors, $\mathcal{O}_{T\tau}$, and make use of Wigner-Eckart theorem, 
\begin{equation}
\langle A_{\Sigma_A}|\mathcal{O}_{T\tau}|B_{\Sigma_B}\rangle = 
\frac{C_{S_B\Sigma_B,T\tau}^{S_A\Sigma_A}}{\sqrt{2S_A+1}}\,\langle A\|\mathcal{O}_T\|B\rangle,
\end{equation}
where $C_{a\alpha,b\beta}^{c\gamma}$ are Clebsch-Gordan coefficients and $\langle A \|\mathcal{O}_T\|B\rangle$ is a reduced matrix element~\cite{Varshalovich1988}.
To treat spin tensors $\mathcal{O}_{T\tau}$ consistently, it is necessary to adopt the convention on the phase imparted to these operators by the ladder spin operators,
\begin{equation}
\begin{split}\label{eq:PhaseConventionSphericalTensors1}
&[\hat{S}_{\pm},\mathcal{O}_{T\tau}]=\sqrt{T(T+1)-\tau(\tau\pm1)}\,\mathcal{O}_{T\tau\pm1},\\
&[\hat{S}_z,\mathcal{O}_{T\tau}]=\tau\,\mathcal{O}_{T\tau},
\end{split}
\end{equation}
where the rising and lowering operators $\hat{S}_{\pm}=\hat{S}_x\pm i \hat{S}_y$ are related to the spherical tensor component of spin through $\hat{S}_{\pm}=\mp\sqrt{2}S_{\pm1}$. The convention in~\eqref{eq:PhaseConventionSphericalTensors1} is consistent with~\cite{Helgaker} (see Eq.~2.3.1), as well as with~\cite{Varshalovich1988} (see Eq.~1 in ch.~3).

For each orbital $p$, the two creation operators $a^\dagger_{p_\pi}$, with $\pi=\pm\frac{1}{2}$, comply with the phase conventions~\eqref{eq:PhaseConventionSphericalTensors1}, whereas the adjoint operators $a_{p_\pi}\equiv(a_{p_\pi}^\dagger)^\dagger$ do not. For this reason, it is convenient to define a second set of tensors operators $b^\dagger_{p_\pi}$ and $b_{p_\pi}$,
\begin{equation}
b^\dagger_{p_\pi} = a^\dagger_{p_\pi},\qquad b_{p_\pi} = (-1)^{\frac{1}{2}+\pi} a_{p_{-\pi}},
\end{equation}
which satisfy~\eqref{eq:PhaseConventionSphericalTensors1}, as well as the convention
\begin{equation}\label{eq:PhaseConventionSphericalTensors2}
(b_{p_\pi})^\dagger = (-1)^{\frac{1}{2}+\pi}b^\dagger_{p_{-\pi}}.
\end{equation}
Two spherical tensors $A_{T\tau}$ and $B_{J\mu}$ can be coupled to a rank-$K$ tensor as
\begin{equation}
[A_T\otimes B_J]_{K\kappa} \equiv\sum_{\tau\mu}C_{T\tau,J\mu}^{K\kappa}\,A_{T\tau}B_{J\mu}.
\end{equation}
Conversely, the product of two spherical tensors can be expanded in spin-coupled products
\begin{equation}
A_{T\tau}B_{J\mu} = \sum_{K\kappa}C_{T\tau,J\mu}^{K\kappa} [A_T\otimes B_J]_{K\kappa}.
\end{equation}
In particular,
\begin{equation}
\begin{split}
a^\dagger_Pa^\dagger_Q&=a^\dagger_{p_\pi}a^\dagger_{q_\theta}=b^\dagger_{p_\pi}b^\dagger_{q_\theta}=\sum_{T\tau}C_{\frac{1}{2}\pi,\frac{1}{2}\theta}^{T\tau} \,[b^\dagger_{p}\otimes b^\dagger_{q}]_{T\tau},\\
a^\dagger_Pa^\dagger_Qa^\dagger_R&=\sum_{J\mu T\tau}C_{\frac{1}{2}\pi,T\tau}^{J\mu}C_{\frac{1}{2}\theta,\frac{1}{2}\rho}^{T\tau} \,[b^\dagger_{p}\otimes [ b^\dagger_{q}\otimes b^\dagger_r]_{T}]_{J\mu}.
\end{split}
\end{equation}
These expressions, when used in combination with Wigner-Eckart's theorem, allow us to compute matrix elements for arbitrary values of $n$ spin magnetic quantum numbers from selected matrix elements in which two of those $n$ quantum numbers are fixed,
\begin{equation}
\langle A \| \mathcal{O}_{T} \| B\rangle = \frac{\sqrt{2S_A+1}}{C_{S_BM,T0}^{S_AM}}\,\,\langle A_M | \mathcal{O}_{T0} | B_M\rangle.
\end{equation}
This step is essential for the current \ASTRA{} implementation, which relies on quantum-chemistry calculations that consider a single well-defined spin-projection $M$ at a time.
For tensors with an equal number of creation and annihilation operators, with $T=0,\,1,\,2$, it is always possible to find an $M$ such that $C_{S_AM,T0}^{S_BM}\neq 0$. 
Notice that
\begin{equation}\label{eq:Combb}
\begin{split}
[b_p\otimes b_q]_{T\tau}&=(-1)^{1-T}[b_q\otimes b_p]_{T\tau},\\ [b^\dagger_p\otimes b^\dagger_q]_{T\tau}&=(-1)^{1-T}[b^\dagger_q\otimes b^\dagger_p]_{T\tau}.
\end{split}
\end{equation}

\section{Transition Density Matrices}\label{app:TDM}
\ASTRA{} evaluates matrix elements within a CC space obtained by augmenting a given set of multiconfiguration ionic states with arbitrary single-electron functions. To this end, it is convenient to define the first, second, and third order transition-density matrices (TDMs) as the matrix elements between ionic states of strings containing one, two, or three creator operators followed by an equal number of annihilation operators (normal order),
\begin{eqnarray}
\rho^{BA}_{QP}&=& \langle A | a^\dagger_P a_Q |B\rangle,\\
\pi^{BA}_{RS,PQ}&=& \langle A | a^\dagger_P a^\dagger_Q a_S a_R|B\rangle,\\
\gamma^{BA}_{STU,PQR}&=& \langle A | a^\dagger_P a^\dagger_Q a^\dagger_R a_U a_T a_S |B\rangle.
\end{eqnarray}
The rest of this section describes the coupling scheme used in \ASTRA{} to express these TDMs in terms of their reduced counterpart.

\subsubsection{Inactive, active, and virtual orbitals.}

In the expressions that involve TDMs between ionic states, it is convenient to distinguish between: i) inactive spinorbitals, which are identically represented in all of the ions, and which we designate with the letters $W$, $X$, $Y$, $Z$; ii) active orbitals, i.e., those without a well defined occupation in the CI of the ion; and, iii) virtual orbitals, which lie outside the active space, and hence have zero occupation in all the ions. Any TDM element with virtual-orbitals indexes, of course, is zero. The matrix elements with inactive orbitals, which also often vanish, admit simple expressions in terms of lower-order density matrices. In particular,
\begin{equation}\nonumber
\rho^{BA}_{QX} =  \rho^{BA}_{XQ} =\delta_{AB}\delta_{XQ},
\end{equation}
\begin{equation}\nonumber
\begin{split}
&\pi^{BA}_{XS,PQ} =  \pi^{BA}_{SX,QP} = \\
&\quad=(1-\delta_{XS})(1-\delta_{PQ})\left(\delta_{PX}\rho^{BA}_{SQ}-\delta_{QX}\rho^{BA}_{SP}\right),\\
&\pi^{BA}_{RS,PX} =  \pi^{BA}_{SR,XP} = \\
&\quad=(1-\delta_{RS})(1-\delta_{PX})\left(\delta_{SX}\rho^{BA}_{RP}-\delta_{RX}\rho^{BA}_{SP}\right),\\
&\pi^{BA}_{RS,XY} =  \pi^{BA}_{XY,SR} = \\
&\quad=\delta_{AB}(1-\delta_{RS})(1-\delta_{XY})(\delta_{SY}\delta_{RX}-\delta_{SX}\delta_{RY}),\\
&\pi^{BA}_{XP,YQ} =  -\pi^{BA}_{XP,QY} = \\
&\quad=(1-\delta_{XP})(1-\delta_{YQ})(\delta_{XY}\rho^{BA}_{PQ}-\delta_{QX}\delta_{PY}\delta_{AB}),\\
&\pi^{BA}_{XY,ZQ} =  \delta_{AB}(1-\delta_{XY})(1-\delta_{ZQ})\times\\
&\quad\times(\delta_{XZ}\delta_{QY}-\delta_{QX}\delta_{YZ}).
\end{split}
\end{equation}
These formulas allow us to incorporate the two-body interactions with the electrons in the core orbitals in effective one-body operators.

\subsubsection{Reduced Transition Density Matrices}
The quantum chemistry code \LUCIA{} does not provide arbitrary TDM elements, since it restricts the total spin projection to be the same for the two ions. However, within this constraint, \LUCIA{} does provide a sufficient number of \emph{independent} matrix elements to reconstruct all the others by means of the Wigner Eckart theorem (WE). To take advantage of this theorem, it is necessary to cast the string of operators involved in the TDM in terms of well-defined spin tensors. As discussed in the previous section, the first step to use creators and destructor operators consistently with the phase conventions for spherical tensors, we have to express them in terms of the $b$ and $b^\dagger$ operators. Subsequently, we can apply WE to express any TDM elements in terms of reduced matrix elements of its tensorial components. 

In the case of the one-body TDM,
\begin{eqnarray}
\rho^{BA}_{QP} &=& (-1)^{\frac{1}{2}-\theta} \langle A | b^\dagger_{p\pi} b_{q-\theta} |B\rangle=\\
&=&(-1)^{\frac{1}{2}-\theta}\sum_{T\tau}C_{\frac{1}{2}\pi,\frac{1}{2}-\theta}^{T\tau} \langle A | [b^\dagger_p\otimes b_q]_{T\tau}|B\rangle=\nonumber\\
&=& \frac{(-1)^{\frac{1}{2}-\theta}}{\Pi_{S_A}}\sum_{T}\mathsf{R}^{BA}_{[q,p]_T}\sum_{\tau}C_{\frac{1}{2}\pi,\frac{1}{2}-\theta}^{T\tau} C_{S_B \Sigma_B,T\tau}^{S_A\Sigma_A}\nonumber
\end{eqnarray}
where $\Pi_{ab\cdots}=\sqrt{(2a+1)(2b+1)\cdots}$ and we have introduced the notation $\mathsf{R}^{BA}_{[q,p]_T}$ to designate the reduced one-body TDM,
\begin{equation}
\mathsf{R}^{BA}_{[q,p]_T}=\langle A \| [b^\dagger_p\otimes b_q]_T\|B\rangle.
\end{equation}
To evaluate the reduced one-body TDMs from selected matrix elements between states with the same spin projection, let's consider the cases $T=0,1$ separately. For $T=0$, the matrix $\mathsf{R}^{BA}_{[q,p]_{0}}$ is non zero only if the two ions have the same multiplicity. For any value of their projection $\Sigma_A=\Sigma_B$, we can compute the reduced matrix element as 
\begin{equation}
\begin{split}
\mathsf{R}^{BA}_{[q,p]_{0}}&=\sqrt{S_A+1/2}\,\sum_{\sigma}\rho^{BA}_{q_\sigma p_\sigma}.
\end{split}
\end{equation}
For $T=1$, the reduced matrix element is zero if both $A$ and $B$ are singlet states. For parent ions with an odd number of electrons, it is possible to compute $\mathsf{R}^{BA}_{[q,p]_{1}}$ from the uncoupled density matrix between the states with $\Sigma_A=\Sigma_B=\Sigma=1/2$,
\begin{equation}
\begin{split}
\mathsf{R}^{BA}_{[q,p]_{1}} &= \frac{\sqrt{S_A+1/2}}{C_{S_B 1/2,10}^{S_A1/2}} ( \rho^{BA}_{q_\alpha p_\alpha} - \rho^{BA}_{q_\beta p_\beta}).
\end{split}
\end{equation}
For parent ions with an even number of electrons, the matrix element between ions with different spin can be computed choosing $\Sigma_A=\Sigma_B=\Sigma=0$ (which is required if one of the two spin is zero). For parent ions with the same, non-zero spin, however, it is necessary to use any other spin projection (e.g., $\Sigma_A=\Sigma_B=\Sigma=1$)
\begin{equation}
\begin{split}
\mathsf{R}^{BA}_{[q,p]_{1}} &= \frac{\sqrt{S_A+1/2}}{C_{S_B \Sigma,10}^{S_A\Sigma}} ( \rho^{BA}_{q_\alpha p_\alpha} - \rho^{BA}_{q_\beta p_\beta}).
\end{split}
\end{equation}

In the two-body TDM, we first couple the spin of the two creation operators and those of the two annihilation operators to a total spin $T=0,\,1$ and $J=0,\,1$, respectively. These are subsequently coupled to a total spin $K=0,\,1,\,2$,
\begin{eqnarray}
&&\pi^{BA}_{RS,PQ}=-(-1)^{-\sigma-\rho} \langle A | b^\dagger_{p\pi} b^\dagger_{q\theta} b_{s_{-\sigma}} b_{r_{-\rho}}|B\rangle =\nonumber\\
&&=-(-1)^{-\sigma-\rho} \Pi_{S_A}^{-1}\sum_{JTK} \mathsf{\Pi}^{BA}_{[[sr]_J,[pq]_T]_K}\,\times\\
&&\times\,\sum_{\mu\tau\kappa}C_{S_B\Sigma_B,K\kappa}^{S_A\Sigma_A}C_{T\tau,J-\mu}^{K\kappa}C_{\frac{1}{2}-\sigma,\frac{1}{2}-\rho}^{J-\mu}C_{\frac{1}{2}\pi,\frac{1}{2}\theta}^{T\tau}\nonumber
\end{eqnarray}
where the reduced matrix element $\mathsf{\Pi}^{BA}_{[[sr]_J,[pq]_T]_K}$ is
\begin{equation}\nonumber
\mathsf{\Pi}^{BA}_{[[sr]_J,[pq]_T]_K} = \langle A \| [[b^\dagger_p\otimes b^\dagger_q]_T\otimes[b_s\otimes b_r]_J]_K\|B\rangle.
\end{equation}
As shown in \eqref{eq:Combb}, spin-coupled electron pairs anti-commute or commute depending on whether they form a singlet or a triplet, respectively, which implies the following permutation symmetries,
\begin{equation}\label{eq:ComPi}
\begin{split}
\mathsf{\Pi}^{BA}_{[[sr]_J,[pq]_T]_K} &= (-1)^{J+1} \mathsf{\Pi}^{BA}_{[[rs]_J,[pq]_T]_K}= \\
&=(-1)^{T+1}\mathsf{\Pi}^{BA}_{[[sr]_J,[qp]_T]_K}=\\
&=(-1)^{J+T} \mathsf{\Pi}^{BA}_{[[rs]_J,[qp]_T]_K}.
\end{split}
\end{equation}
The two-body reduced matrix elements have the following expression in terms of TDMs between states with the same spin projection $\Sigma_A=\Sigma_B=\Sigma$,
\begin{equation}\nonumber
\begin{split}
\mathsf{\Pi}^{BA}_{[[sr]_J,[pq]_{T}]_K}&=\frac{(-1)^{J} \Pi_{S_A}}{C_{S_B\Sigma,K0}^{S_A\Sigma}}\sum_{\pi\theta\sigma\rho\tau}(-1)^{\tau}\,\times\\
&\times\,C_{T\tau,J-\tau}^{K0}C_{\frac{1}{2}\sigma,\frac{1}{2}\rho}^{J\tau}C_{\frac{1}{2}\pi,\frac{1}{2}\theta}^{T\tau}\pi^{BA}_{r_\rho s_\sigma,p_\pi q_\theta}.
\end{split}
\end{equation}
Finally, the coupling scheme for the 3B-TDM is 
\begin{eqnarray}
\gamma^{BA}_{STU,PQR}&
=&\zeta \langle A | b^\dagger_{p\pi} b^\dagger_{q\theta} b^\dagger_{r\rho} b_{u_{-\upsilon}} b_{t_{-\tau}} b_{s_{-\sigma}} |B\rangle=\nonumber\\
&=&\frac{\zeta}{\Pi_{S_A}} \sum_{CDFJK} \mathsf{\Gamma}^{BA}_{[[[ut]_Fs]_J,[[pq]_Cr]_D]_K} \,\,\times\nonumber\\
&\times&\sum_{\gamma \delta \varphi \mu \kappa} 
C_{\frac{1}{2}\pi,\frac{1}{2}\theta}^{C\gamma} C_{C\gamma,\frac{1}{2}\rho}^{D\delta}
C_{\frac{1}{2}-\upsilon,\frac{1}{2}-\tau}^{F\varphi}\times\nonumber\\
&\times& C_{F\varphi,\frac{1}{2}-\sigma}^{J\mu}
C_{D\delta,J\mu}^{K\kappa} C_{S_B\Sigma_B,K\kappa}^{S_A\Sigma_A}
\end{eqnarray}
where $\zeta=(-1)^{-1/2-\upsilon-\tau-\sigma}$ and
\begin{eqnarray}
&&\mathsf{\Gamma}^{BA}_{[ [ [ut]_F s]_J,[[pq]_Cr]_D]_K} =\\
&&=\langle A \|\, [[[b^\dagger_{p}\otimes b^\dagger_{q}]_C\otimes b^\dagger_{r}]_D \otimes[[b_{u}\otimes b_{t}]_F\otimes b_{s}]_J]_K \|B\rangle.\nonumber
\end{eqnarray}

%%%%%%%%%%%%%%%%%%%%%%%%%%%%%%%%%%%%%%%%%%%%%%%%%%%%%%%%%%%%%%%%%%%%%%%%%%%%
\section{Matrix elements between CC states \label{app:MatrixElements}}
This appendix derives the overlap as well as the matrix elements of one-body and two-body operators between CC states, without and with spin adaptation. The evaluation of each of the relevant matrix elements between CC states starts with the reduction of the corresponding operator string.

\subsection{Overlap}
The CC states are generally not orthogonal, so it is necessary to evaluate their overlap,
\begin{equation}\label{Overlap}
\langle A,P|B,Q\rangle = \langle  A | a_P a^\dagger_Q|B\rangle.
\end{equation}
In normal form, the string of operators is 
\begin{equation}\label{Ap:strOv}
a_P a_Q^\dagger =  S_{P Q}-a^{\dagger}_{Q}a_{P},
\end{equation}
where $S_{PQ}=\langle P | Q \rangle$, and hence
\begin{equation}\label{Overlap}
\langle A,P|B,Q\rangle = \langle  A | a_P a^\dagger_Q|B\rangle=S_{PQ} \delta_{AB}- \rho^{BA}_{PQ}.
\end{equation}
The CC states built from internal active orbitals, therefore, in general are not normalized,
\begin{equation}
\langle A,P|A,P\rangle = 1 - \rho^{AA}_{PP}.
\end{equation}
Using \eqref{Overlap} and \eqref{eq:coupledstate}, we obtain the expression for the overlap between SACC states,
\[\begin{split}
&\langle A,p|B,q\rangle =\\
&=\sum_{\Sigma_A\Sigma_B\pi\theta}\hspace{-8pt}C_{S_A \Sigma_A,\frac{1}{2}\pi}^{S\Sigma}C_{S_B \Sigma_B,\frac{1}{2}\theta}^{S\Sigma} \left(S_{pq}\delta_{\pi,\theta}\delta_{AB}- \rho^{BA}_{p_{\pi}q_{\theta}}\right)=\\
&=\delta_{AB}S_{pq}-\sum_{\Sigma_A\Sigma_B\pi\theta}C_{S_A \Sigma_A,\frac{1}{2}\pi}^{S\Sigma}C_{S_B \Sigma_B,\frac{1}{2}\theta}^{S\Sigma} \rho^{BA}_{p_{\pi}q_{\theta}}.
\end{split}\]
Summing the Clebsch-Gordan coefficients, we can rewrite the overlap as 
\[
\langle A,p|B,q\rangle =\delta_{AB}S_{pq}+W_{pq}^{BA},
\]
where
\[
W^{BA}_{pq}=\eta\sum_T\mathsf{R}^{BA}_{[p,q]_T} \Pi_{T}\sjs{T}{1/2}{1/2}{S}{S_B}{S_A},
\]
and $\eta = (-1)^{S_B-1/2+S}$.

\subsection{One-body operators}
The matrix element between CC states of a generic spin-free one-body operator is 
\begin{equation}
\langle A ,P| \hat{O} |B,Q\rangle = o_{RS} \langle A | a_P a^\dagger_R a_S a^\dagger_Q |B\rangle,
\end{equation}
where summation over repeated magnetic quantum numbers and orbital indexes is implied. In normal form, the operator string reads
\begin{equation}\label{Ap:str1B}
\begin{split}
a_P a_R^\dagger a_S a_Q^\dagger = \delta_{P R}\delta_{Q S}+S_{P Q}a^{\dagger}_{R}a_{S}-\delta_{P R}a^{\dagger}_{Q}a_{S}\\
-\delta_{Q S}a^{\dagger}_{R}a_{P}+a^{\dagger}_{Q}a^{\dagger}_{R}a_{P}a_{S},
\end{split}
\end{equation}
and hence the matrix element becomes
\[\begin{split}
\label{eq:obocc}
\langle A,P | \hat{O} |B,Q\rangle &= o_{PQ}\,\delta_{AB}+S_{PQ} \,O_{AB}-\\
&- o_{PS}\,\rho^{BA}_{S,Q}-\rho^{BA}_{P,R}\,o_{RQ} +
o_{RS}\, \pi^{BA}_{SP,QR}
\end{split}\]
where $O_{AB}=\langle A|\hat{O}|B\rangle$. Notice that we can evaluate $O_{AB}$ within \ASTRA{} starting from the one-body TDM provided by \LUCIA{}.
Such one-body spin-free operator vanish unless $\Sigma_A=\Sigma_B$
\begin{equation}
\langle A | \hat{O} |B\rangle = o_{rs} \langle A | a^\dagger_{r\sigma}a_{s\sigma}  |B\rangle = o_{rs}\rho^{BA}_{s_{\sigma}r_{\sigma}}.
\end{equation}
Since the two spin projections coincide, the expression for $\rho$ becomes
\begin{equation}\nonumber\begin{split}
\rho^{BA}_{s_{\sigma}r_{\sigma}} &= \sum_{T}\frac{\mathsf{R}^{BA}_{[s,r]_T}}{\Pi_{S_A}}\sum_{\tau\sigma}(-1)^{\frac{1}{2}-\sigma}C_{\frac{1}{2}\sigma,\frac{1}{2}-\sigma}^{T\tau} C_{S \Sigma,T\tau}^{S\Sigma}=\\
&=\sum_{T}\frac{C_{S \Sigma,T0}^{S\Sigma}\mathsf{R}^{BA}_{[s,r]_T}}{\Pi_{S_A}}\sum_\sigma (-1)^{\frac{1}{2}-\sigma}C_{\frac{1}{2}\sigma,\frac{1}{2}-\sigma}^{T0} =\\
&=\frac{\mathsf{R}^{BA}_{[s,r]_0}}{\sqrt{2}\,\Pi_{S_A}}.
\end{split}\end{equation}
We introduce here the quantity
\begin{equation}\label{Eq:Qtensor}
Q^{BA}_{sr}=\frac{\sqrt{2}\,\mathsf{R}^{BA}_{[s,r]_0}}{\Pi_{S_A}}
\end{equation}
with which
\begin{equation}
\langle A | \hat{O} |B\rangle = o_{rs} \frac{\Pi^{-1}_{S_A}}{\sqrt{2}}\mathsf{R}^{BA}_{[s,r]_0}=o_{rs}Q^{BA}_{sr}/2.
\end{equation}
The extra factor of $2$ in the definion of $Q^{BA}_{rs}$ is the legacy of an arbitrary convention. As shown further down, $Q^{BA}_{rs}$ is containd also in the expression of two-body operators, where it appears without prefactors.

It is now possible to obtain the expression for the matrix element between SACC states. After some straightforward recoupling of tensor operators and contraction of Clebsch-Gordan coefficients, one finds
\begin{equation}
\begin{split}
&\langle A,p| \hat{O} |B,q\rangle =
S_{pq}\,O_{AB} +o_{pq}\,\delta_{AB}  +\\
&+\eta\sum_T\left(\mathsf{R}^{BA}_{[p,r]_T}o_{rq}+o_{ps}\mathsf{R}^{BA}_{[s,q]_T}\right)\,\Pi_{T}\times\\
&\qquad\times\sjs{T}{1/2}{1/2}{S}{S_B}{S_A}+\\
&+\eta\, o_{rs} \sum_{KTJ}(-1)^{J+K}\,\mathsf{\Pi}^{BA}_{[[ps]_J,[qr]_T]_K}\,\Pi_{KTJ}\times\\
&\qquad\times\,\sjs{S_A}{K}{S_B}{1/2}{S}{1/2} \sjs{J}{T}{K}{1/2}{1/2}{1/2}.
\end{split}
\end{equation}
We can condense the notation by using the tensor $W_{pq}^{BA}$ introduced above, and the new tensor
\begin{equation}\begin{split}
P^{BA}_{ps,qr}&=\eta\sum_{KTJ}(-1)^{J+K}\mathsf{\Pi}^{BA}_{[[ps]_J,[qr]_T]_K}\,\Pi_{KTJ}\times\\
&\times\sjs{S_A}{K}{S_B}{1/2}{S}{1/2} \sjs{J}{T}{K}{1/2}{1/2}{1/2} ,
\end{split}\end{equation}
with which the matrix element becomes
\begin{equation}
\begin{split}
&\langle A,p| \hat{O} |B,q\rangle =
S_{pq}\,O_{AB} +o_{pq}\,\delta_{AB} +\\
&+W^{BA}_{pr}\,o_{rq} + o_{ps}\,W^{BA}_{sq}+o_{rs}\,P^{BA}_{ps,qr}. 
\end{split}
\end{equation}

The orbitals used to augment the ions do not include inactive orbitals. As a consequence, the summation over the orbital indexes in the third and fourth term can be restricted to the active orbitals only. On the other hand, inactive orbitals do give a contribution to the fifth term, which can be expressed in terms of one-body density matrices. By explicitly indicating the summation over orbitals indexes, and labelling with a prime the sums over active orbitals only, the result reads 
\begin{eqnarray}
&&\langle A,p| \hat{O} |B,q\rangle =
S_{pq}\,O_{AB} +o_{pq}\,\delta_{AB} +\langle O \rangle_{\mathrm{core}}\,W^{BA}_{pq}+\nonumber\\
&&+{\sum_r}'\left(W^{BA}_{pr}o_{rq} + o_{pr}W^{BA}_{rq}\right)+{\sum_{rs}}'o_{rs}P^{BA}_{ps,qr},\label{eq:O1b}
\end{eqnarray}
where $\langle O \rangle_{\mathrm{core}}=2\sum_xo_{xx}$.  Notice that only when both ions are augmented by an active internal orbital do all terms in the expression contribute to the matrix element. If one of the two orbitals is external, only one of the terms involving the $B$ tensors survive. If both orbitals are external, only the first two terms survive.

\subsection{Two-body operators}
The matrix element of a two-body spin-free operator between CC states is 
\begin{equation}
\langle A,P|\hat{G}|B,Q\rangle= \frac{1}{2}[TU|RS]\langle A | a_P a^\dagger_{T}a^\dagger_{R}a_{S}a_{U}a^\dagger_{Q}|B\rangle.
\end{equation}
A term on the right-hand side vanish, of course, unless the spin projections of $T$ and $U$, and of $R$ and $S$ coincide. For the sake of clarity, however, we do not split the orbital and spin indexes yet. In normal form, the operator string reads
\begin{eqnarray}\label{Ap:str2B}
\begin{split}
& a_P a_T^\dagger a_R^\dagger a_S a_U a_Q^\dagger =\\
&=\delta_{P R}\delta_{Q S}\,a^{\dagger}_{T}a_{U}-\delta_{U Q}\delta_{P R}\,a^{\dagger}_{T}a_{S}-\\
&-\delta_{T P}\delta_{Q S}\,a^{\dagger}_{R}a_{U}+\delta_{T P}\delta_{U Q}\,a^{\dagger}_{R}a_{S}+\\
&+\delta_{P R}\,a^{\dagger}_{T}a^{\dagger}_{Q}a_{U}a_{S}+\delta_{Q S}\,a^{\dagger}_{T}a^{\dagger}_{R}a_{U}a_{P}-\\
&-\delta_{P Q}\,a^{\dagger}_{T}a^{\dagger}_{R}a_{U}a_{S}+\delta_{U Q}\,a^{\dagger}_{T}a^{\dagger}_{R}a_{P}a_{S}+\\
&+\delta_{T P}\,a^{\dagger}_{Q}a^{\dagger}_{R}a_{U}a_{S}+a^{\dagger}_{T}a^{\dagger}_{Q}a^{\dagger}_{R}a_{U}a_{P}a_{S}.
\end{split}
\end{eqnarray}
The general expression for the matrix element, considering that the operator is spin free, becomes
\begin{equation}\label{eq:tbocc2}
\begin{split}
&\langle A,P|\hat{G}|B,Q\rangle=
G_{AB} S_{PQ} +\\
&+\delta_{\pi\theta}[pq|rs]\rho^{BA}_{s_\rho,r_\rho}-[ps|rq]\rho^{BA}_{s_\pi,r_\theta}+\\
&+[pt|rs]\pi^{BA}_{t_\pi s_\rho,r_\rho q_\theta} + [qt|rs]\pi^{BA}_{p_\pi  s_\rho,r_\rho t_\theta} -\\
&-\frac{1}{2}[tu|rs]\gamma^{BA}_{u_\tau s_\rho p_\pi,t_\tau r_\rho q_\theta}.
\end{split}
\end{equation}
We can now obtain an expression for the matrix element of the two-body operator between SACC states. The full expression of the result is not only lenghty to derive but also rather long. Luckily, as in the case of the overlap and of the one-body operators, it is possible to express it in condensed form with the help of \emph{ad hoc} auxiliary tensors. We have already defined three of these tensors, $Q^{BA}_{sr}$, $W^{BA}_{rs}$, and $P^{BA}_{rs,tu}$. 
It is now convenient to define
\begin{eqnarray}
&&C^{BA}_{psu,trq} = 
\frac{\eta}{2}\sum_{CDFJK} 
(-1)^{K+D+C-1/2}\,\Pi_{CDFJK}\times\nonumber\\
&&\times\sjs{S_A}{K}{S_B}{1/2}{S}{1/2}\sjs{J}{C}{1/2}{1/2}{K}{D}\sjs{C}{1/2}{J}{F}{1/2}{1/2}\times\nonumber\\
&&\times\,\,\mathsf{\Gamma}^{BA}_{[[[ps]_Fu]_J,[[tr]_C q]_D]_K}.
\end{eqnarray}
After making the necessary contractions, the total electron-electron repulsion matrix element between SACC states can be rewritten as
\begin{eqnarray}
\langle A,p| \hat{G} |B,q\rangle &=&
S_{pq} \,G_{AB} +\nonumber\\
&+&[pq|rs]\,Q^{BA}_{sr} +[ps|rq]\, W^{BA}_{sr} +\nonumber\\
&+&[pt|rs]\, P^{BA}_{ts,qr} +[qt|rs]\,P^{BA}_{ps,tr}+\nonumber\\
&+&[tu|rs]\,  C^{BA}_{psu,trq}.
\end{eqnarray}
Notice that the last term involving the third-order TDM needs to be evaluated exclusively when both ions are augmented by an active ionic orbital. Since in this case \emph{all} the orbitals involved are GTOs, the matrix element is most conveniently evaluated within the \LUCIA{} Quantum Chemistry code, which also computes the TDMs. 

\subsection{Hamiltonian matrix elements\label{app:MatrixElements:Hamiltonian}}
As in the case of the mono-electronic operators, it is convenient to factor out the contribution of the inactive orbitals in the two-body matrix element. In doing so, it will emerge that the core orbitals contribute to the inter-electronic repulsion as an effective one-body potential. The most natural context to separate effective one-body and two-body terms is in the evaluation of the full Hamiltonian, which we presently proceed to carry out.

Here, we assume that the Hamiltonian is diagonal in the ion basis, $\langle A | H | B \rangle = E_A\delta_{AB}$. The matrix element of the total Hamiltonian between SACC states then becomes
\begin{equation}\label{eq:twobody}
\begin{split}
&\langle A,p | \hat{H} |B,q\rangle=\\
&=\delta_{A B} (S_{pq} E_A + h_{pq})+\\
&+W^{BA}_{pr}h_{rq} + h_{ps}W^{BA}_{sq}+h_{rs}P^{BA}_{ps,qr}+\\
&+[pq|rs]Q^{BA}_{sr} +[ps|rq] W^{BA}_{sr} +\\
&+[pt|rs] P^{BA}_{ts,qr} + [qt|rs]P^{BA}_{ps,tr}+\\
&+[tu|rs]  C^{BA}_{psu,trq}.
\end{split}
\end{equation}
All the summations over the ionic orbitals can be restricted to the active orbitals only, provided that the mono-electronic Hamiltonian $h_{pq}$ is replaced by an effective Hamiltonian $\tilde{h}_{pq}$ that incorporates the Coulomb and exchange terms with the core,
\begin{equation}
\tilde{h}_{pq}=h_{pq} + 2J^{\mathrm{core}}_{pq}-K^{\mathrm{core}}_{pq},
\end{equation}
where
\begin{equation}
J^{\mathrm{core}}_{pq}=\sum_x^{\mathrm{core}} [pq|xx],\quad
K^{\mathrm{core}}_{pq}=\sum_x^{\mathrm{core}}[px|xq].
\end{equation}
The total matrix element of the Hamiltonian reads
\begin{eqnarray}
&&\langle A,p | \hat{H} |B,q\rangle =\nonumber\\
&&=\delta_{A B} (S_{pq} E_A + \tilde{h}_{pq})
\,+\,E_{\mathrm{core}}\,W^{BA}_{pq}\,+\nonumber\\
&&+{\sum_r}'\Big(W^{BA}_{pr}\tilde{h}_{rq} +\tilde{h}_{pr}W^{BA}_{rq}\Big)+{\sum_{rs}}'\tilde{h}_{rs}P^{BA}_{ps,qr}\nonumber\\
&&+{\sum_{rs}}'\Big([pq|rs]Q^{BA}_{sr}+[ps|rq] W^{BA}_{sr}\Big)+\label{eq:Hll}\\
&&+{\sum_{rst}}'\Big([pt|rs] P^{BA}_{ts,qr} + [qt|rs]P^{BA}_{ps,tr}\Big)+\nonumber\\
&&+{\sum_{turs}}'[tu|rs]  C^{BA}_{psu,trq}.\nonumber
\end{eqnarray}
Notice that in this formulation, the matrix elements lend themselves to an efficient implementation.
The Hamiltonian matrix elements between active channels can be computed by the \LUCIA{} program. In \ASTRA{}, therefore, \eqref{eq:Hll} is used only when at least one of the two augmented orbital-indexes, $p$ or $q$, does not correspond to an active internal orbital. Let's assume, therefore, without loss of generality, that $q$ is a virtual orbital (the case in which $p$ is virtual and $q$ isn't is obtained by Hermitian conjugation). The expression then greatly simplifies to
\begin{eqnarray}
&&\langle A,p | \hat{H} |B,q\rangle =\delta_{A B} (S_{pq} E_A +\tilde{h}_{pq})+\nonumber\\
&&+{\sum_r}'W^{BA}_{pr}\tilde{h}_{rq}+{\sum_{rs}}'\Big([pq|rs]Q^{BA}_{sr}+[ps|rq] W^{BA}_{sr}\Big)+\nonumber\\
&&+{\sum_{rst}}'[qt|rs]P^{BA}_{ps,tr}\qquad\qquad q\in\mathrm{VO}.\label{eq:H_AMO_VHO}
\end{eqnarray}
If both $p$ and $q$ are virtual orbitals, the formula simplifies further to
\begin{eqnarray}
&&\langle A,p | \hat{H} |B,q\rangle =\delta_{A B} (S_{pq} E_A +\tilde{h}_{pq})+\nonumber\\
&&+{\sum_{rs}}'\Big([pq|rs]Q^{BA}_{sr}+[ps|rq] W^{BA}_{sr}\Big).\label{eq:Hpaqv}
\end{eqnarray}
Owing to the compact support character of the spherical numerical functions used to describe the $N-$th electron at large distances, and to the rapid decrease with $r$ of any of the GTO active orbitals used to build the ions, for all practical purposes, the product of most of the former functions with \emph{any} of the latter is zero. The virtual orbitals with this property are effectively external to the molecular region. If \emph{either} $p$ or $q$ is an external spherical orbital (ESO), then the expression for the Hamiltonian simplifies even further. For example, if $q$ is an ESO,
\begin{eqnarray}
&&\langle A,p | \hat{H} |B,q\rangle =\delta_{A B} (S_{pq} E_A +\tilde{h}_{pq})+\nonumber\\
&&+{\sum_{rs}}'[pq|rs]Q^{BA}_{sr}\qquad\qquad q\in\mathrm{ESO}.\label{eq:Hpvqv}
\end{eqnarray}
Finally, if $q$ is an AMO and $p$ is an ESO, their supports are effectively disjoint and hence the Hamiltonian matrix element vanishes entirely.

\subsubsection{Matrix elements between AC blocks.}
The evaluation with the TDMCC formalism of the Hamiltonian matrix elements between ionic states that are both augmented by active orbitals requires the use of 3-body TDMs, which is both computationally and logistically demanding. For this reason, it is convenient to use auxiliary matrix elements that are generated by the same code that prepares the TDMs. \LUCIA{} can generate elements of the form
\begin{equation}
\langle A| a_P \hat{H}a^\dagger_Q |B\rangle,
\end{equation}
in which the ion and electron spins are not coupled. Once a sufficient number of such independent matrix elements are available, it is possible to extract the reduced matrix elements of the associated singlet and triplet operators $[b_p\hat{H}b_q^\dagger]_{T\tau}$, and from these evaluate those between SACC states. Let's define
\begin{equation}\begin{split}
&\mathsf{H}^{AB}_{[p,q]_T}\equiv(-1)^{T+1}\langle A\| [b_p \hat{H}b^\dagger_q]_T \|B\rangle =(-1)^{T+1}\Pi_{S_A}^{-1} \times\\
&\times\hspace{-8pt}\sum_{\Sigma_A\Sigma_B\pi\theta\tau}\hspace{-8pt}(-1)^{\frac{1}{2}-\pi} C_{\frac{1}{2}-\pi,\frac{1}{2}\theta}^{T\tau}C_{S_B\Sigma_B,T\tau}^{S_A\Sigma_A} \langle A| a_P \hat{H}a^\dagger_Q |B\rangle.\nonumber
\end{split}\end{equation}
The reduced element $\mathsf{H}^{AB}_{[p,q]_T}$ can be determined from the matrix elements between states with a same (suitable) $\Sigma=\Sigma_A=\Sigma_B$, and operators with the same spin projection, $\pi=\theta$,
\begin{equation}
\mathsf{H}^{AB}_{[p,q]_T}= \sum_{\theta}\frac{\Pi_{S_A} C_{\frac{1}{2}\theta,\frac{1}{2}-\theta}^{T0}}{(-1)^{\frac{1}{2}-\theta}C_{S_B\Sigma,T0}^{S_A\Sigma}} \langle A| a_P \hat{H}a^\dagger_Q |B\rangle,\nonumber
\end{equation}
Notice that this is the same relation as the one between $W^{BA}_{[q,p]_T}$ and $\rho^{BA}_{q_\theta p_\theta}$. Once the matrix elements $\mathsf{H}^{AB}_{[p,q]_T}$ are available, the Hamiltonian between SACC states can be shown to be given by
\begin{equation}
\langle A,p | \hat{H} |B,q\rangle
=-\eta\sum_T\mathsf{H}^{AB}_{[p,q]_T} \Pi_{T}\sjs{T}{1/2}{1/2}{S}{S_B}{S_A}.\nonumber
\end{equation}

\section{CIS Model}\label{app:CIS}

\ASTRA{} was tested for consistency by comparison with a CIS model, in which the configuration space comprises a single determinant for the neutral ground state, $|\mathbf{K}\rangle$, as well as all the single excitations $|\mathbf{K}_A^P\rangle = a_P^\dagger a_A |\mathbf{K}\rangle$ ($P\neq A$, $\langle P|Q\rangle=\delta_{PQ}$ $\forall P,\,Q$). This space, of course, coincides with the CC space generated by all the ions $|A\rangle= |\mathbf{K}_A\rangle \equiv a_A|\mathbf{K}\rangle$.
All CIS states are ortonormal,
\begin{equation}
\langle \mathbf{K}_A^P|\mathbf{K}_B^Q\rangle = \delta_{PQ}\delta_{AB},\quad \langle \mathbf{K}| \mathbf{K} \rangle = 1,\quad \langle \mathbf{K}_A^P|\mathbf{K}\rangle = 0.\nonumber
\end{equation}
The matrix elements of arbitrary one-body and two-body operators between such CIS states are readily determined with Slater rules~\cite{SzaboOstlund}. For monoelectronic operators,
\begin{equation}\begin{split}
&\langle \mathbf{K} | \hat{\mathcal{O}} |\mathbf{K} \rangle = \sum_{I\in\mathbf{K} } o_{II}=\langle\hat{\mathcal{O}}\rangle_{\mathbf{K}},\quad
\langle \mathbf{K} | \hat{\mathcal{O}} |\mathbf{K}_B^Q \rangle = o_{AQ},\\
&\langle  \mathbf{K}_A^P | \hat{\mathcal{O}} | \mathbf{K}_B^Q\rangle = \delta_{AB}\left[\delta_{PQ}\left(\langle\hat{\mathcal{O}}\rangle_{\mathbf{K}} -o_{AA}\right) + o_{PQ} \right].\nonumber
\end{split}\end{equation}

Let us introduce the notation $\langle IJ||KL\rangle = \langle IJ|KL\rangle - \langle IJ|LK\rangle=[IK|JL]-[IL|JK]$. For the matrix elements between CIS states of the inter-electronic repulsion then are
\begin{equation}\begin{split}
\langle \mathbf{K} | &\hat{G} |\mathbf{K} \rangle = \frac{1}{2}\sum_{I,J\in\mathbf{K} } \langle IJ\|IJ\rangle,\\
\langle \mathbf{K} | &\hat{G}  |\mathbf{K}_B^Q \rangle = \sum_{I\in\mathbf{K}}\langle BI\|QI\rangle,\\
\langle  \mathbf{K}_A^P |&\hat{G}  | \mathbf{K}_A^P\rangle = \frac{1}{2}\sum_{I,J\in\mathbf{K}_A^P } \langle IJ\|IJ\rangle,
\end{split}\nonumber
\end{equation}

\begin{equation}\begin{split}
\langle  \mathbf{K}_A^P |&\hat{G}  | \mathbf{K}_A^{Q\neq P}\rangle = \sum_{I\in\mathbf{K}}  \langle AI\|QI\rangle + \langle AP\|QP\rangle,\\
\langle  \mathbf{K}_A^P |&\hat{G}  | \mathbf{K}_{B\neq A}^P\rangle = \sum_{I\in\mathbf{K}}  \langle BI\|AI\rangle + \langle BP\|AP\rangle,\\
\langle  \mathbf{K}_A^P |&\hat{G}  | \mathbf{K}_{B\neq A}^{Q\neq P}\rangle =\langle BP\|AQ\rangle.
\end{split}\nonumber
\end{equation} 
If the initial state is a closed-shell Hartree-Fock state, then the expressions can be simplified into sums over the doubly occupied orbitals. These results are sufficient to implement an \emph{ad hoc} code to test the accuracy of the results obtained with the more sophisticated \ASTRA{} code, in the CIS limit.

Thanks to its simplicity, the CIS approximation offers also a way of testing the consistency of the phase conventions in the interface between \LUCIA{} and \ASTRA{}. Indeed, within the CIS approximation, the TDM have the following elementary expressions
\[\begin{split}
\rho^{BA}_{QP}&=\langle \mathbf{K}|a_A^\dagger a_P^\dagger a_Q a_B|\mathbf{K}\rangle=\nonumber\\
&=(1-\delta_{AP})\,(1-\delta_{BQ})\,(\delta_{AB}\delta_{PQ}-\delta_{AQ}\delta_{PB}),
\end{split}\]
\[\begin{split}
&\pi^{BA}_{RS,PQ}=\langle \mathbf{K}|a_A^\dagger a^\dagger_P a^\dagger_Q a_S a_R a_B|\mathbf{K}\rangle=\\
&=(1-\delta_{AP})(1-\delta_{AQ})(1-\delta_{PQ})\times\\
&\times(1-\delta_{BR})(1-\delta_{BS})(1-\delta_{RS})\times\notag \\
&\left(\delta_{AB}\delta_{PR}\delta_{QS}-\delta_{AB}\delta_{QR}\delta_{PS}\ +\delta_{QB}\delta_{AR}\delta_{PS}\right.\notag\\
&\left.-\delta_{PB}\delta_{AR}\delta_{QS}+\delta_{PB}\delta_{QR}\delta_{AS}-\delta_{QB}\delta_{PR}\delta_{AS}\right).\notag
\end{split}\]

%\bibliography{ASTRA3}


\end{document}